\documentclass[a4paper,fleqn,usenatbib,useAMS]{mnras}

\usepackage{graphicx}	
\usepackage{tabularx}
\usepackage{amsmath}	
\usepackage{amssymb}	
\usepackage{multicol}        
\usepackage[percent]{overpic}

\usepackage[T1]{fontenc}
\usepackage{ae,aecompl}

\title[GENESIS: Self-Consistent Models of Exoplanetary Spectra]{GENESIS: New Self-Consistent Models of Exoplanetary Spectra}

\author[Gandhi \& Madhusudhan]{Siddharth Gandhi$^{1}$\thanks{Contact e-mail: \href{mailto:sng29@ast.cam.ac.uk,nmadhu@ast.cam.ac.uk}{sng29@ast.cam.ac.uk, nmadhu@ast.cam.ac.uk}} \& Nikku Madhusudhan$^{1}$
\\
$^{1}$Institute of Astronomy, University of Cambridge, Madingley Road, Cambridge, CB3 0HA, UK}

\date{Last updated XX; in original form 2016 December 15}

\pubyear{2016}

\begin{document}
\label{firstpage}
\pagerange{\pageref{firstpage}--\pageref{lastpage}}
\maketitle

\begin{abstract}
We are entering the era of high-precision and high-resolution spectroscopy of exoplanets. Such observations herald the need for robust self-consistent spectral models of exoplanetary atmospheres to investigate intricate atmospheric processes and to make observable predictions. Spectral models of plane-parallel exoplanetary atmospheres exist, mostly adapted from other astrophysical applications, with different levels of sophistication and accuracy. There is a growing need for a new generation of models custom-built for exoplanets and incorporating state-of-the-art numerical methods and opacities. The present work is a step in this direction. Here we introduce GENESIS, a plane-parallel, self-consistent, line-by-line exoplanetary atmospheric modelling code which includes (a) formal solution of radiative transfer using the Feautrier method, (b) radiative-convective equilibrium with temperature correction based on the Rybicki linearisation scheme, (c) latest absorption cross sections, and (d) internal flux and external irradiation, under the assumptions of hydrostatic equilibrium, local thermodynamic equilibrium and thermochemical equilibrium. We demonstrate the code here with cloud-free models of giant exoplanetary atmospheres over a range of equilibrium temperatures, metallicities, C/O ratios and spanning non-irradiated and irradiated planets, with and without thermal inversions. We provide the community with theoretical emergent spectra and pressure-temperature profiles over this range, along with those for several known hot Jupiters. The code can generate self-consistent spectra at high resolution and has the potential to be integrated into general circulation and non-equilibrium chemistry models as it is optimised for efficiency and convergence. GENESIS paves the way for high-fidelity remote sensing of exoplanetary atmospheres at high resolution with current and upcoming observations. 
\end{abstract}




\section{Introduction}
\label{sec:introduction}

\begin{table*}
\begin{tabularx}{\textwidth}{p{1.8cm}p{2.5cm}p{2.3cm}p{2.0cm}l*{4}{c}cp{1cm}}
\hline \\
\textbf{Model} & \textbf{R-T Solver$^1$} & \textbf{T-correction$^2$} & \textbf{Parent code} & \textbf{Opacity} & \textbf{Convection}  & \textbf{Clouds} & \textbf{Scattering}\\ \\
\hline \\ 
Seager et al. 1998, 2005 & Feautrier & Entropy \newline conservation & UMA  & line-by-line & Yes & Yes & Yes  \\ \\
Barman et al. 2001, 2005 & Short~characteristics+ALI & Uns{\"o}ld-Lucy \newline correction & PHOENIX & line-by-line & Yes & Yes &Yes \\ \\
Fortney et al. 2006, 2008& 2-stream source function & Linearise~flux \newline transfer& McKay~et~al. \newline 1989  & corr.-k & Yes & Yes & Yes\\ \\
Burrows et al. 2008 &  Feautrier/DFE+ALI  & Rybicki & TLUSTY  & line-by-line & Yes & Yes & Yes \\ \\
Molli\`{e}re et al. 2015, 2017& Feautrier &Variable Eddington factors & New & corr.-k & Yes & Yes & Yes \\ \\
Drummond et al. 2016 & ALI & Local flux \newline balance & New  & corr.-k  & Yes & Yes & Yes\\ \\
Malik et al. 2017 & 2-stream approximation & Local flux \newline balance & New  & corr.-k  & No & Yes & Yes\\ \\
\textbf{This work} & Feautrier & Rybicki & New & line-by-line & Yes& No & Yes$^\dagger$\\ 
\hline
\end{tabularx}
\caption{Comparison of prominent atmospheric models in the literature including strong irradiation. Details of the methods used by Seager et al. can be found in \citet{seager_code} with methods in \citet{seager_code_method},  \citet{Barman_code} describe their methods in \citet{Barman_method}, \citet{fortney_2007,fortney_2008} in \citet{Fortney_method} and \citet{mckay_1989}, Burrows et al. in \citet{Sudarsky_2003} with their methods in \citet{hubenybook}, Malik et al. in \citet{Malik_2016} and \citet{heng_2014}, Molli\`{e}re et al. in \citet{molliere_code} and \citet{Molliere2016}, and \citet{drummond_2016} in \citet{amundsen_2014}. \newline 
$^1$ Radiative Transfer solver used in the model. \newline
$^2$ Temperature correction scheme used to ensure radiative-convective equilibrium. \newline 
The remaining columns describe other aspects, including whether convection and clouds are incorporated in the model, and whether the opacities are treated in a line-by-line or using the correlated-k approximation. See section~\ref{sec:introduction} for a discussion of the different models. \newline
$^\dagger$In the present work we only consider weak scattering, due to gaseous H$_2$ Rayleigh scattering.
}\label{comp_table}
\end{table*}

The study of exoplanetary atmospheres is a major frontier in exoplanetary science. Numerous observational surveys in the last decade have led to the detection of thousands of exoplanets and have revealed that exoplanets are extremely common and extremely diverse in their masses, radii, and orbital architectures \citep{fischer2016}. New surveys in the near future, such as the TESS, CHEOPS, and PLATO space missions and several ground-based surveys, are expected to find thousands more planets orbiting nearby stars. The proximity of these host stars would allow detailed characterisation of their  planetary atmospheres. Exoplanetary science is now entering an exciting time as low mass planets are already being discovered in the habitable zones of nearby stars \citep{gillon2016,escude2016}, thereby opening the possibility of detecting biosignatures in their atmospheres in the future. 

There has been tremendous progress in the observations and characterisation of exoplanetary atmospheres in recent years \citep{Madhu_review2014,heng2015,crossfield2016,Madhu_review2016}. For transiting exoplanets, a wealth of new high quality  transmission and emission spectra have been observed thanks to the HST WFC3 spectrograph as well as large ground-based facilities. These and other observations have led to the detections of atomic and molecular species, clouds/hazes, thermal inversions, day-night circulation patterns, and exospheres \citep[e.g.,][]{deming2013,mccullough2014,Madhu_2014,kreidberg2014,stevenson2014,ehrenreich2015, Wyttenbach2015,Sing2016}. These observations are a substantial improvement over the previous years when largely broadband photometric or low-resolution observations were available \citep[see e.g., review by][]{Madhu_review2014}. At the same time, for directly imaged planets ground-based high resolution spectra of several young giant planets have led to molecular detections in their atmospheres \citep[e.g.,][]{barman2015,macintosh2015}. On the other hand, molecular detections in the atmospheres of transiting and non-transiting exoplanets have also been made using high dispersion spectroscopy with very high resolution (R$\sim10^5$) spectra in the near infrared \citep{snellen2010,brogi2012,birkby2013, wyttenbach_2017}. These different means of characterising exoplanetary atmospheres are all poised to receive a major boost with 
upcoming large facilities, such as the James Webb Space Telescope (JWST) and the large ground-based telescopes (e.g., E-ELT, TMT, etc). 

Central to the characterisation of exoplanetary atmospheres is the availability of high fidelity spectral models. Two modelling approaches have been developed over time to address complementary needs: forward models and retrieval methods \citep[see e.g., reviews by ][]{Madhu_review2014,heng2015,Madhu_review2016}. Forward models attempt to self-consistently model detailed physicochemical processes in the atmospheres under various assumptions for chemical abundances, energy transport mechanisms, chemical equilibrium/non-equilibrium, etc. Such models are extremely useful for a priori theoretical understanding of atmospheric processes in exoplanets under varied conditions, for predicting observables to aid in planning observations, and for initial interpretation of data. A wide range of such models have been developed since the first atmospheric observations \citep[e.g.,][]{seager_code}, ranging from 1-D spectral models discussed below to detailed models of atmospheric dynamics, non-equilibrium chemistry, and other aspects of exoplanetary atmospheres. 1-D spectral models are also useful to incorporate into 3-D general circulation models \citep{showman_gcm,kataria2015} as well as into non-equilibrium chemistry models \citep{moses2013} to predict their spectral signatures. 

On the other hand, retrieval methods are inverse modelling techniques which attempt to formally fit models to spectral data to derive statistically robust constraints on the atmospheric properties through the constrained model parameters. This approach, which followed the first infrared spectra of exoplanets \citep[e.g.][]{Madhu_retrieval2009,madhu_2011_nature}, involves combining 1-D parametric models with as few assumptions as possible with statistical parameter estimation techniques to efficiently explore the model parameter space. Retrieval methods have been the workhorse for deriving statistical estimates of atmospheric chemical abundances and temperature profiles from exoplanetary spectra in recent years \citep[e.g.][]{lee2012,line2013,Madhu_2014,benneke2015,waldmann2015,lavie2016}.

Both forward models (or `self-consistent' models) and retrieval methods are essential for thorough characterisation of exoplanetary atmospheres. The constraints on atmospheric properties derived from observations using retrieval methods need to be checked against self-consistent forward models to understand the conformance or deviations of the constrained solutions with respect to the assumptions of self-consistent models. The differences, if any, could lead to refinement of the models, constraints on non-equilibrium phenomena, and/or to discover new physicochemical effects unaccounted for in the forward models. Efficient forward models of exoplanetary spectra are therefore vital to gain a good insight into exoplanetary atmospheres, particularly of those quite hostile worlds stretching our current knowledge at the extremes of atmospheric conditions well beyond those encountered in the solar system. 

Over the past two decades various groups have developed self-consistent models with different levels of complexity and accuracy. Here we discuss some of the prominent plane-parallel self-consistent forward models of atmospheric spectra in the field that take into account strong irradiation, as relevant for currently known transiting exoplanets \citep{seager_code, Sudarsky_2003, Barman_code, fortney_2007, fortney_2008, burrows_2008,molliere_code,Molliere2016,Malik_2016,drummond_2016}. Table~\ref{comp_table} shows a comparison of these models, many of which have been adapted from pre-existing forward models originally built to model radiative transfer in stellar atmospheres \citep[e.g.,][]{seager_code_method,Hauschildt_phoenix,Hubeny_tlusty}, atmospheres of planets in the solar system \citep{marley_uranus} or circumstellar discs \citep{molliere_method}. Typically, such models assume a given chemical composition and solve for the temperature profile and emergent spectrum of the atmosphere under assumptions of radiative-convective equilibrium, hydrostatic equilibrium, and local thermodynamic equilibrium (LTE), for given system parameters and boundary conditions. Typically, models assume thermochemical equilibrium to determine the chemical composition for assumed elemental abundances. 

\begin{table*}
\begin{tabular}{|r l l l l| l|}
  \hline
  \multicolumn{5}{|c|}{\textbf{Inputs}} & \textbf{Key Features}\\
  \textbf{Stellar Properties:} & $R_{\mathrm{star}}$ & $\mathrm{log}(g_{\mathrm{star}})$ & $T_{\mathrm{eff,star}}$ & $Z_{\mathrm{star}}$ & RTE using Feautrier Method\\ 
  \textbf{Planetary Properties:} &  $R_{\mathrm{planet}}$ & $\mathrm{log} (g_{\mathrm{planet}})$ & $T_{\mathrm{int,planet}}$ & $a_{\mathrm{planet}}$ & Rybicki's method for temperature iterator\\ 
   & d &  $f_{\mathrm{r}}$ & Mixing length $l$ & \{X/H\} & Chemical Equilibrium option\\
  \cline{1-5}
   \multicolumn{5}{|c|}{\textbf{Outputs}} & Convection with mixing length theory\\
  \multicolumn{2}{|c}{Spectrum} &\multicolumn{3}{c|}{Profiles of $P$, $T$, $\rho$, chemical species} & Line-by-line opacities\\
  \cline{1-5}
  \textbf{Not yet included:}& \multicolumn{3}{c}{Clouds} & &Irradiated and non-irradiated atmospheres\\
  \hline
\end{tabular}
\caption{The inputs, outputs and key features of GENESIS. $R_{\mathrm{star}}$, $\mathrm{log}(g_{\mathrm{star}})$, $T_{\mathrm{eff,star}}$ and $Z_{\mathrm{star}}$ are the radius, $\log({\rm gravity})$, metallicity, and effective temperature of the planet hosting star. $R_{\mathrm{planet}}$, $\mathrm{log}(g_{\mathrm{planet}})$, $T_{\mathrm{int,planet}}$ and $a_{\mathrm{planet}}$ are the radius, $\log({\rm gravity})$, internal temperature, and the orbital separation of the planet. $d$ is the distance to the system. The stellar redistribution factor $f_\mathrm{r}$ denotes the fraction of the insolation received by the dayside atmosphere, accounting for spherical geometry, day-night energy redistribution, etc. The mixing length for the convection $l$ (usually taken to be the scale height) and the elemental abundances relative to atomic hydrogen \{X/H\} are also inputs. The model has the option to either fix the molecular abundances of the main gaseous species at equilibrium values or set them to other specified values for each layer. The outputs are the emergent spectrum, either the planet-star flux ratio or the planet flux alone, the pressure-temperature (P-T) profile and the chemical profiles. Clouds/hazes have not been included in the present work.}
\label{tab:params}
\end{table*}

The differences between the various forward models in Table~\ref{comp_table} lie in the sophistication of their numerical methods and assumptions therein. Some key aspects where differences lie include the approaches for solving the radiative transfer equation (RTE) and for ensuring radiative-convective equilibrium, the opacities used (e.g. line-by-line vs correlated-k approximation), and the incorporation of physical processes such as clouds, convection, etc. Models adapted from pre-existing stellar atmosphere codes \citep[e.g.,][]{Barman_paper,burrows_2008}, such as TLUSTY or PHOENIX, have the advantage of very accurate radiative transfer solvers and temperature correction procedures. On the other hand, such models are based on pre-computed opacity grids with assumed chemical compositions, e.g. of chemical equilibrium with solar abundances, and hence less flexible/efficient to explore a wide range in chemical parameter space. On the other hand, codes adapted from planetary applications \citep[][]{fortney_2008,marley2012} assume the two-stream source function approximation to solve the RTE, which is computationally efficient but the implementation is first order accurate (see e.g. \citealt{hubeny_2017,hubenybook}).

More recent models have been custom-built for exoplanetary atmospheres. \citet{molliere_code} and \citet{Molliere2016}  developed a model based on the methods of \citet{molliere_method}, originally developed for radiative transfer in protoplanetary discs. They solve the RTE using the Feautrier method and determine the temperature profile in equilibrium  using the variable Eddington factor approach, and include scattering due to condensate species along with opacities computed using the k-distribution method. \citet{drummond_2016} have constructed a model that uses accelerated lambda iteration for their radiative transfer, and also including scattering from condensate species \citep{ackerman_2001} and opacities using the correlated-k approximation. Another recently developed code, HELIOS \citep{Malik_2016}, adopts the two-stream approximation for the radiative transfer, and also the correlated-k approximation for opacities, albeit with differences in temperature correction methods and treatment of scattering. \cite{Malik_2016} have implemented an analytic solution of the transfer equation that allows one to specify an arbitrary number of streams in the limit of pure absorption. A detailed comparison of these different models can be found in \cite{hubeny_2017}. These recent codes \citep{drummond_2016, Molliere2016,Malik_2016} are customised to explore a wide range of chemical compositions (e.g. metallicities and C/O ratios) motivated by recent suggestions of the importance of these quantities \citep{Madhuco_2012,moses2013}.

There is a need for a new generation of models that build upon the past successes to address the new wave of current and upcoming high quality data. The most desirable spectral code arguably is one that obtains an accurate formal solution of the radiative transfer equation (e.g. using the Feautrier method) and robustly derives the temperature correction for radiative-convective equilibrium, e.g. using a formal Rybicki procedure or accelerated Lambda Iteration \citep[see e.g.][]{hubeny_2017}, and considers high density opacity sampling or ``line-by-line" opacities, with flexible chemistry. Such a model is highly desirable for the planning and interpretation of high-precision and high-resolution spectroscopic observations expected from current and upcoming large facilities. On the other hand, such a code would also be invaluable to be integrated into detailed higher dimensional models of non-equilibrium chemistry and general circulation models to understand chemical and physical processes in exoplanetary atmospheres at high resolution. Finally, there is great value to a coherently developed model architecture where all the components, from radiative transfer, energy balance, and chemistry, to line-by-line broadened opacities, are developed with the latest computing practices and languages, and all tuned to exoplanet conditions. Our present work is a step in this direction. 

In this work we report a new state-of-the-art forward model of exoplanetary atmospheres. We introduce GENESIS, a plane-parallel, line-by-line, self-consistent exoplanetary atmospheric modelling code built upon accurate numerical techniques and the latest opacities and chemical resources. The radiative transfer is solved using Feautrier's method, which is second-order accurate, and the radiative-convective equilibrium is established using Rybicki's method with complete linearisation \citep{hubenybook}. The molecular opacity is calculated from the most modern line lists available to compute the latest high-temperature cross sections with accurate prescriptions for temperature and pressure broadening. The chemical abundances are derived assuming thermochemical equilibrium with variable elemental abundances, in order to facilitate efficient exploration of the chemical phase space (spanning C/O ratios, metallicities, visible absorbers, etc.). The code self-consistently and simultaneously treats both incident irradiation and internal flux so that it is applicable to both irradiated exoplanets as well as isolated planets. 

In order to demonstrate the GENESIS code in the present work we focus on cloud-free giant exoplanets, both irradiated and  non-irradiated, and explore a range of model parameters. All components of the code are extensively tested for accuracy and convergence and provide good match to published models. We generate high-resolution emergent spectra and pressure-temperature (P-T) profiles for models over a wide range in equilibrium temperature (i.e., degree of irradiation), metallicity, C/O ratio and internal flux, spanning both planets with and without strong irradiation. Additionally, for planets with strong irradiation we explore the dependence of the models on the visible absorption, via Na/K and TiO opacity. In particular, we demonstrate how the strength of thermal inversions depend on the TiO abundance. We also report model spectra and P-T profiles for several known hot Jupiters.

In what follows we first describe in section~\ref{sec:methods} the various components of the GENESIS code and the numerical methods used to model the atmosphere. We present the model grid and results in section~\ref{sec:results}, including models of several known systems. We present a summary of the results and a discussion of ongoing and future work in section~\ref{sec:summary}. 

\section{Methods}
\label{sec:methods}


\begin{figure}


\includegraphics[width=0.8\columnwidth]{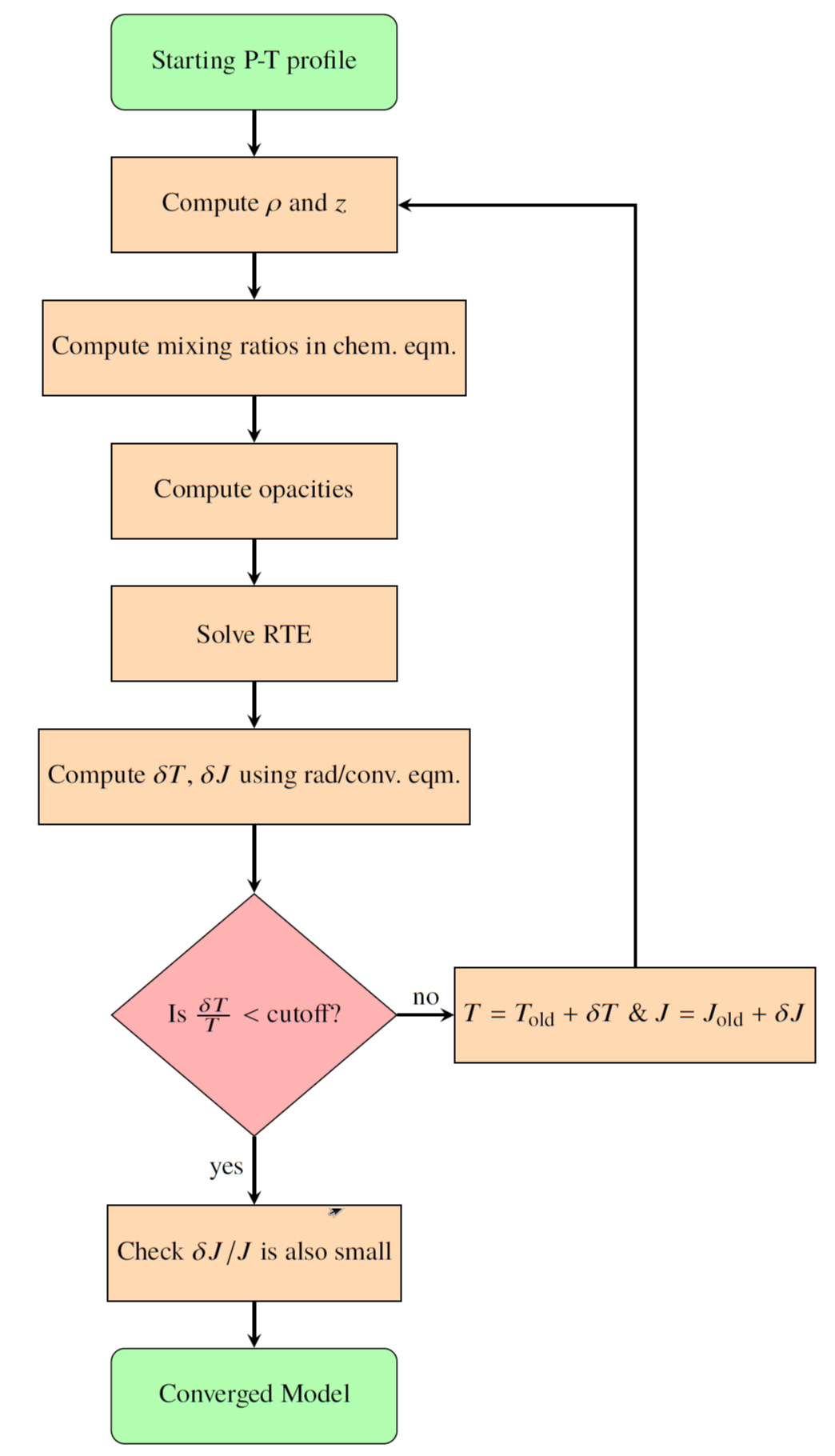}
\caption{The GENESIS Modelling Framework. The flowchart shows the key components of the model and the steps followed to converge to radiative-convective equilibrium. The description of each step is given in section~\ref{sec:overview}.}
\label{fig:flowchart}
\end{figure}

We model a fully self-consistent plane parallel atmosphere in radiative-convective equilibrium, local thermodynamic equilibrium and chemical equilibrium. Our goal is to compute model emergent spectra and the atmospheric profiles of temperature, pressure, density, and composition. In the present work, for purposes of demonstrating the model, we focus on cloud-free H$_2$-dominated atmospheres, i.e. of gas giants. Extension to other planetary types can be easily achieved by incorporating the required chemistry and planetary properties. Here we discuss the different components of the model. 

\subsection{Basic Equations}

The main equations governing the model are as follows. 
\begin{align}
\frac{dP}{dz} & = -\rho g, \label{eqn:PT1}\\
P & = \frac{\rho k_b T}{\overline{m}},\label{eqn:idealgas}\\
\mu \frac{dI_\nu}{d\tau_{\nu}} & = I_\nu-S_\nu, \\ 
d\tau_{\nu} & = -(\kappa_{\nu}+\sigma_{\nu}) dz, \label{eqn:tau}\\
\int_0^\infty & \kappa_\nu (J_\nu-B_\nu) d\nu = 0.
\end{align}

The first equation describes the hydrostatic equilibrium for a fluid in one dimension and the second is the ideal gas equation, with the pressure $P$, temperature $T$, mean molecular mass $\overline{m}$, and density $\rho$ in a layer of the atmosphere at a distance $z$ in the vertical direction. The third is the Radiative Transfer Equation (RTE) describing the transfer of photons through an optical depth $\tau_\nu$ at an angle cosine $\mu$ relative to the vertical, with a specific intensity $I_\nu$ and a source function $S_\nu$. $\tau_\nu$ depends on the extinction coefficient, which is the sum of the absorption coefficient $\kappa_\nu$ and scattering coefficient $\sigma_\nu$ (equation \ref{eqn:tau}). $\kappa_\nu$ is the cumulative absorption from all the species. The final equation is the balance of energy entering and leaving a given layer of the atmosphere, the radiative equilibrium equation, with the mean intensity of radiation $J_\nu$ and Plank function $B_\nu$. We note that this equation is corrected for the convective flux where the atmosphere is unstable against convection, as discussed in section~\ref{sec:convection}. An overview of the methods is described below, followed by the sections explaining in detail how the methods are implemented. 

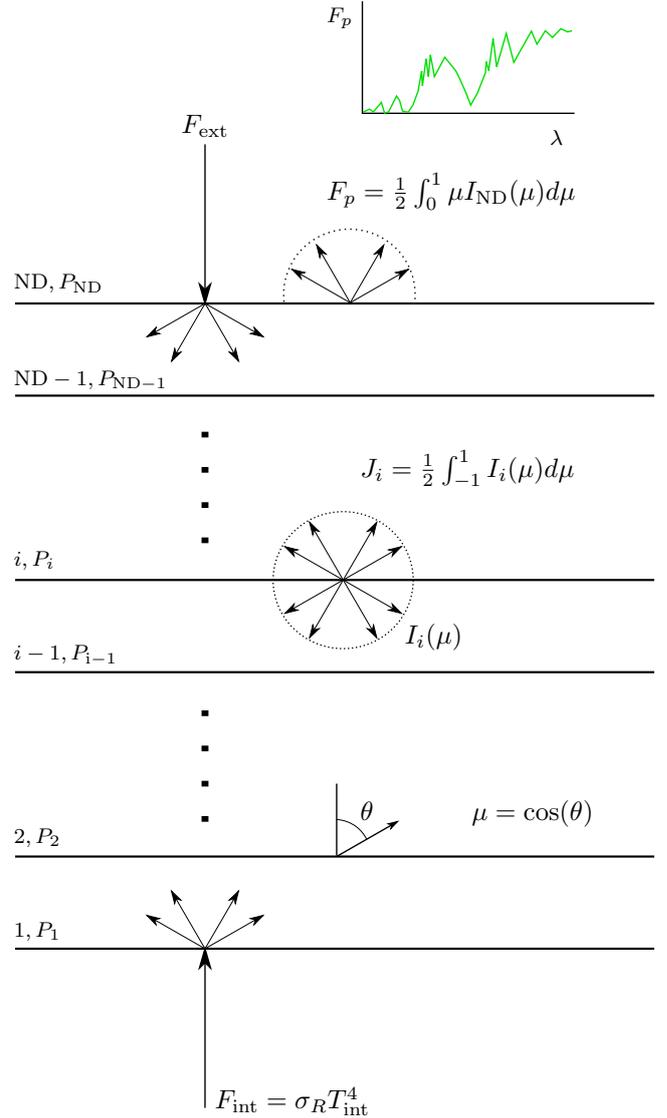
\begin{figure}
\begin{overpic}[width=\columnwidth]{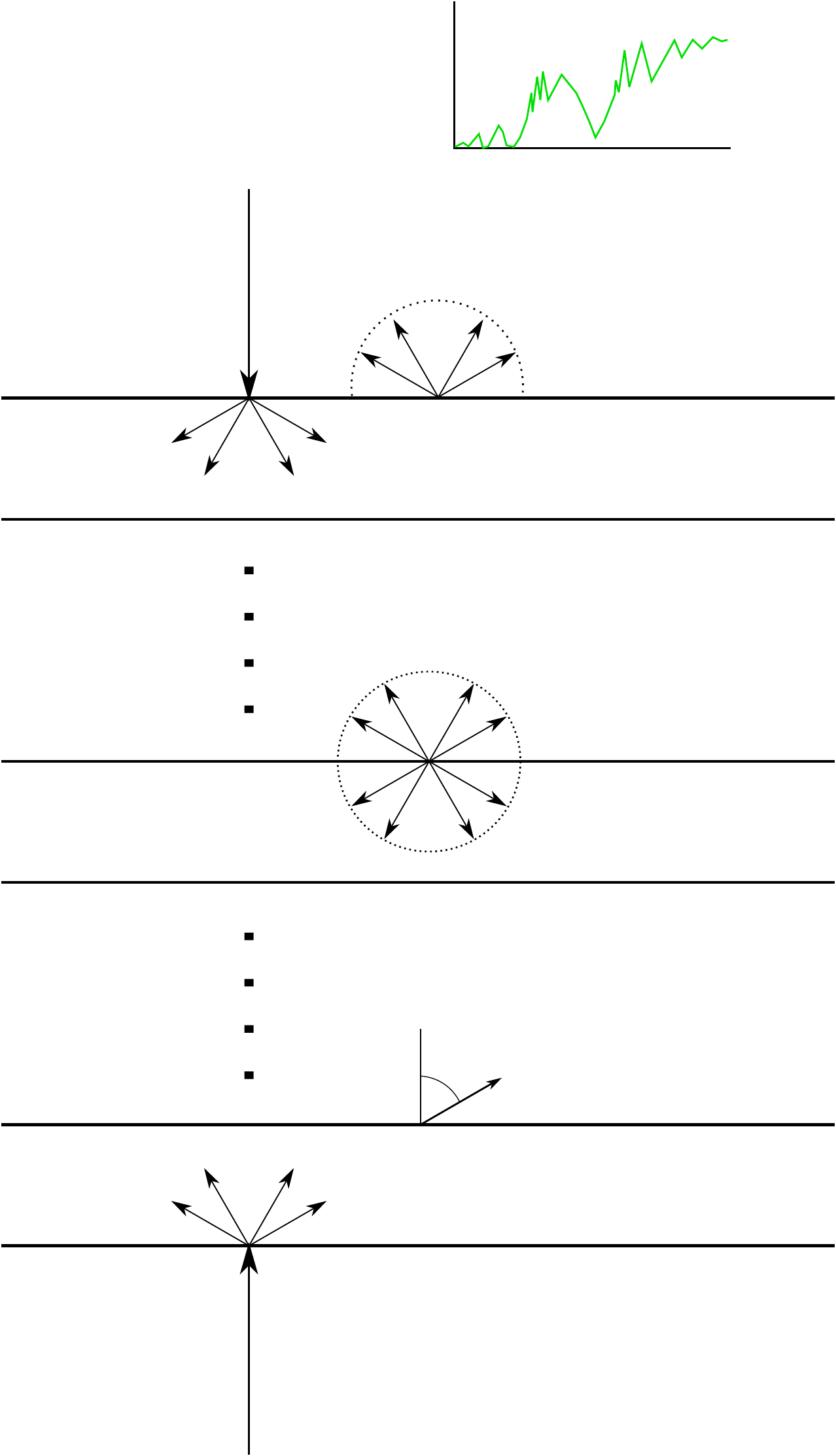}
 \put (48,87) {$\displaystyle\lambda$}
 \put (28,98) {$\displaystyle F_p$}
 \put (28,82) {\large$F_p = \frac{1}{2}\int_{0}^{1} \mu I_{\mathrm{ND}}(\mu) d\mu$}
 \put (18,0) {\large$F_{\mathrm{int}}=\sigma_R T_{\mathrm{int}}^4$}
 \put (35,42) {\large$I_i(\mu)$}
 \put (31,57) {\large$J_i = \frac{1}{2}\int_{-1}^{1} I_i(\mu) d\mu$}
 \put (15,88) {\large$F_{\mathrm{ext}}$}
 \put (31,26) {\large$\theta$}
 \put (41,26) {\large$\mu = \mathrm{cos}(\theta)$}
 \put (0,15.5) {$1,P_1$}
 \put (0,24) {$2, P_2$}
 \put (0,40.5) {$i-1,P_{\mathrm{i-1}}$}
 \put (0,49) {$i,P_i$}
 \put (0,65.3) {$\mathrm{ND-1},P_\mathrm{ND-1}$}
 \put (0,74) {$\mathrm{ND},P_{\mathrm{ND}}$}
\end{overpic}
\caption{Schematic of the model atmosphere. The plane-parallel model comprises of $ND$ layers, with the stellar flux incident at the top ($F_\mathrm{ext}$) and the internal heat flux at the bottom denoted by $F_\mathrm{int}$, corresponding to temperature $T_\mathrm{int}$. The specific intensity is $I$, $J$ is the mean intensity, and $F_p$ is the emergent flux from the top of the atmosphere.}
\label{fig:sketch}
\end{figure}

\subsection{Overview of the Methods}
\label{sec:overview}
Figure~\ref{fig:flowchart} shows a flowchart of the GENESIS model components. Here we give a brief overview of the modelling scheme and elaborate on the details in subsequent sections below. For a fixed pressure grid in the atmosphere, the five unknowns $z$, $T$, $\rho$,  $I_\nu$ and $\tau_\nu$ are fully determined by solving the above five equations. We first begin with a trial pressure-temperature (P-T) profile. The assumption of hydrostatic equilibrium and the ideal gas equation of state are solved to obtain the profiles of $\rho$ and $z$ at each pressure level. The quantities $P$, $T$, and $\rho$, along with the elemental abundances, are then used in a chemical equilibrium module to derive mixing ratios of the major chemical species in the atmosphere. At the same time, $P$ and $T$ are also used to derive the absorption cross sections of the corresponding chemical species. The mixing ratios and the absorption cross sections, along with any scattering, together determine the opacity, and hence the optical depth $\tau_\nu$, in the atmosphere. Then we solve the radiative transfer equation at each sampled frequency to determine the radiation field denoted by the specific intensity $I_\nu$. 

The derived $I_\nu$ for the initial trial $P$-$T$ profile will not necessarily satisfy the radiative-convective equilibrium. Therefore, an adjustment to the P-T profile and the mean intensity is determined using a temperature correction scheme. This procedure is repeated until the temperature is accurate to within a certain tolerance level, which we take to be $10^{-4}$. Once convergence is achieved, the corresponding solution provides the emergent spectrum and the atmospheric profiles. 

The model, therefore, comprises of four key components: (a) Radiative Transfer, (b) Radiative-Convective Equilibrium, (c) Chemical Equilibrium, and (d) Opacities. In what follows, we describe each component of the model in detail. We begin by elaborating on how we solve the radiative transfer equation for a fixed P-T profile (the formal solver) using the Feautrier method, followed by energy transport and how we ensure radiative-convective equilibrium using the Rybicki procedure. We then present the chemical module to determine the chemical abundances, and finally the methods used to determine the opacity, including calculating the absorption cross sections. Finally, we discuss how we account for the stellar flux incident on the planet. 

\subsection{Radiative Transfer}
\label{feautrier method}
To obtain a solution to the RTE for a given P-T profile, we use the Feautrier method described in \cite{mihalasbook}. It is second order accurate, and many angle points can be used without excessive computational expense. It can be modified to exactly solve the transfer equation as well, an advantage in the case of strong scattering (no iterative solving of the RTE is required). The full details of this method and others are explored in detail in \cite{hubenybook}, and for convenience we adopt the notation used in it. The first 3 moments of the specific intensity of radiation $I$ are given by
\begin{align}
J_\nu & = \frac{1}{2}\int_{-1}^{1} I(\mu)d\mu, \\
H_\nu & = \frac{1}{2}\int_{-1}^{1} \mu I(\mu)d\mu,\\
K_\nu & = \frac{1}{2}\int_{-1}^{1} \mu^2 I(\mu)d\mu.
\end{align}
The $J$ term is also known as the mean intensity of radiation. We also define $f_{\nu}$ and $g_{\nu}$ for future convenience
\begin{align}
f_\nu & \equiv K_\nu/J_\nu = \frac{\int_{-1}^{1}I(\mu)\mu^2 d\mu}{\int_{-1}^{1}I(\mu)d\mu},\label{eqn:f_nu} \\
g_{\nu} & \equiv \frac{H_{\nu}(\tau = 0)}{J_{\nu}(\tau = 0)} = \frac{\int_{0}^{1}I(\mu,\tau =0)\mu d\mu}{\int_{-1}^{1}I(\mu,\tau =0)d\mu}.
\end{align}
The source function in the RTE is 
\begin{align}
S_\nu & = \frac{\kappa_\nu B_\nu + \sigma_\nu J_\nu}{\kappa_\nu + \sigma_\nu}.
\end{align}
\begin{align}
B(T,\nu) & = \frac{2h \nu^3}{c^2}\frac{1}{e^\frac{h \nu}{k_b T} - 1},
\end{align}
is the Planck function that describes the spectral radiance of a body at a temperature $T$ and frequency $\nu$.
Now consider 2 beams of radiation travelling in the direction $\pm \mu$. The radiative transfer equations for these are
\begin{align}
\mu \frac{\partial I_\nu(\mu)}{\partial \tau_{\nu}} & =  I_\nu(\mu)-S_\nu(\mu),\label{eqn:rtI1}, \\
-\mu \frac{\partial I_\nu(-\mu)}{\partial \tau_{\nu}} & =  I_\nu(-\mu)-S_\nu(-\mu), \nonumber \\ \label{eqn:rtI2}
& = I_\nu(-\mu)-S_\nu(\mu),
\end{align}
where for the last line we assume that the source function is symmetric in $\mu$. Defining the new quantities 
\begin{align}
j_{\mu,\nu} &= \frac{1}{2}(I_\nu(\mu)+I_\nu(-\mu)), \\
h_{\mu,\nu} &= \frac{1}{2}(I_\nu(\mu)-I_\nu(-\mu)),
\end{align}
then equation \ref{eqn:rtI1} and \ref{eqn:rtI2} can be combined into a second order equation in $j_{\mu,\nu}$.

\begin{align}
\mu^2 \frac{\partial^2 j_{\mu,\nu}}{\partial \tau_{\nu}^2} = j_{\mu,\nu} - S_\nu, \label{eqn:rtj}.
\end{align}
For a given source function $S_\nu$, we can numerically solve for $j_{\mu,\nu}$ to derive the mean intensity $J_\nu$. Any $S(\mu)$ can be used, as long as it satisfies $S(\mu)=S(-\mu)$.
The RTE can also be expressed by an integration over $\mu$ to
\begin{align}
\frac{\partial^2 (f_\nu J_\nu)}{\partial \tau_{\nu}^2} = J_\nu - S_\nu = \frac{\kappa_\nu (J_\nu-B_\nu)}{\kappa_\nu +\sigma_\nu}. \label{eqn:rtJ}
\end{align}
The boundary conditions at the top and bottom of the atmosphere are 
\begin{align}
\left.\frac{\partial (f_\nu J_\nu)}{\partial \tau_\nu}\right|_{\tau = 0} & = g_\nu J_\nu(0) - H_\mathrm{ext}, \label{eqn:topbc}\\
\left.\frac{\partial (f_\nu J_\nu)}{\partial \tau_\nu}\right|_{\tau = \tau_{max}} & = \frac{1}{2}(B_\nu - J_\nu) +\frac{1}{3}\frac{\partial B_\nu}{\partial \tau_\nu}. \label{eqn:bottombc}
\end{align}
equation \ref{eqn:topbc} is a flux conservation condition with external stellar irradiation $H_\mathrm{ext} = F_\mathrm{ext}/(4\pi)$ incident at the top of the atmosphere. The condition at the bottom of the atmosphere (equation \ref{eqn:bottombc}) is known as the diffusion approximation where the atmosphere can be considered to be optically thick due to the strong opacity and gas density, hence photons are diffusive and not ballistic. The current method is unsuitable for strong scattering; iteration of the radiative transfer equation will be required for a fixed P-T profile to converge onto the radiation field if scattering is significant compared to the absorption. In future work the method will be modified to take into account strong scattering and anisotropic scattering due to clouds. The computational method for implementing the solution of the RTE, equation \ref{eqn:rtj}, is described in appendix \ref{appendix:formal_soln}.

\subsection{Radiative-Convective Equilibrium}

The temperature profile in the atmosphere is governed by the energy budget in each layer of the atmosphere. In equilibrium, the temperature profile is required to be such that no net energy accumulates in any layer. In regions of the atmosphere where radiation is the efficient means of energy transport pure radiative equilibrium is satisfied. On the other hand, when the atmosphere is unstable against convection, an additional flux is required to balance the energy flow. Here we discuss the methods we use to ensure radiative-convective equilibrium in our model atmosphere to find a converged temperature profile. 

\subsubsection{Radiative Equilibrium}

The energy flowing into a layer of the atmosphere must equal the energy flowing out if it is in equilibrium. For an initial P-T profile that we begin with, this will not necessarily be the case. To correct the temperature and bring it into equilibrium, we need to know the amount of energy or flux flowing into and out of a layer. Local energy balance between each layer of the atmosphere determines the temperature profile and ensures total global energy conservation. The radiative equilibrium condition in a single layer of the atmosphere can be written as either
\begin{align}
\int_0^\infty \kappa_\nu (J_\nu-B_\nu) d\nu & = 0, \textrm{or} \label{eqn:reqm_int_no_conv}\\ 
\int_0^\infty \frac{d(f_\nu J_\nu)}{d\tau_\nu} d\nu & = \frac{\sigma_R}{4 \pi}T_\mathrm{int}^4.\label{eqn:reqm_diff_no_conv} 
\end{align}
Here, $T_\mathrm{int}$ is referred to as the temperature corresponding to the internal heat flux that emanates from the planet's convective core. Stellar contexts often refer to this as $T_\mathrm{eff}$, but we reserve this notation for the effective temperature of the host star.
Equation \ref{eqn:reqm_int_no_conv} is an energy match condition, and equation \ref{eqn:reqm_diff_no_conv} represents conservation of the total flux entering/leaving in a given layer. Indeed, it may be shown by differentiating equation \ref{eqn:reqm_diff_no_conv} with respect to $\tau$ and using equation \ref{eqn:rtJ} that equation \ref{eqn:reqm_int_no_conv} is identical. The reason why both forms are given is that numerically they both behave differently. Near the top of the atmosphere, equation \ref{eqn:reqm_int_no_conv} is a better constraint to use as the $d\tau$ in equation \ref{eqn:reqm_diff_no_conv} is a small quantity, whereas near the bottom equation \ref{eqn:reqm_int_no_conv} runs into difficulty as numerical instabilities occur when $\kappa$ is large. We apply a switch around the $\sim$1 bar level in the atmosphere to keep the constraint numerically stable.

We use equations \ref{eqn:reqm_int_no_conv} and \ref{eqn:reqm_diff_no_conv} to determine the self-consistent P-T solution iteratively. For a given P-T profile, the equations for hydrostatic equilibrium, the ideal gas equation and the radiative transfer equation are used to give the radiation field. A correction to the temperature is calculated using Rybicki's method by complete linearisation, as we demonstrate in section \ref{Rybicki_method}. This brings the atmosphere into radiative equilibrium. 

Radiative equilibrium may also be enforced using the equation of energy conservation including the full 3-D velocity field of the atmosphere \citep{vallis_2006_book}. 1-D models, including ours, generally set the velocity field to be zero and derive a time independent solution. Full General circulation models can use the 3D energy conservation condition that accounts for the velocity field. Both approaches are equivalent.

\begin{figure}
  \centering
    \includegraphics[trim = 10mm 10mm 25mm 20mm, clip,width=\columnwidth]{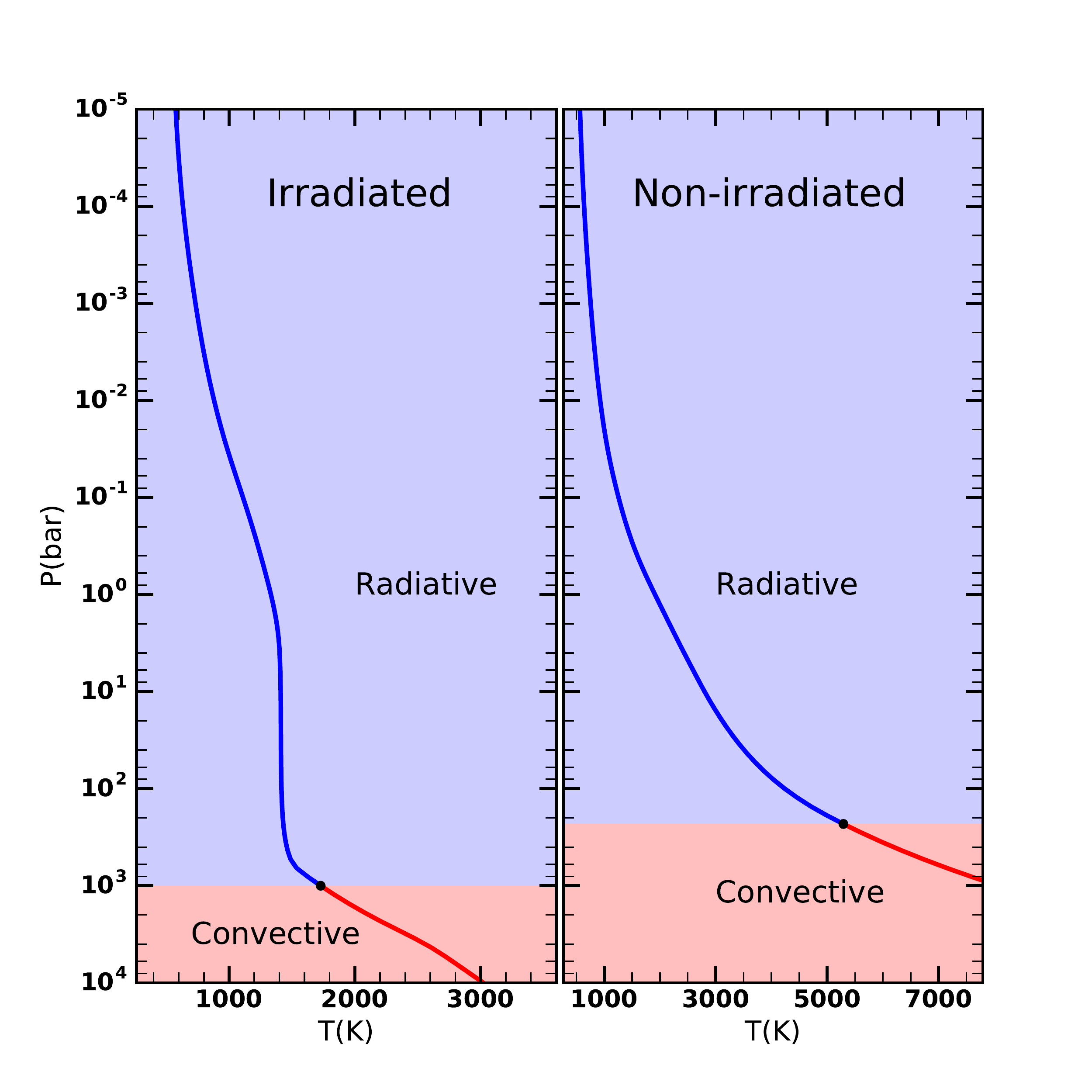}
  \caption{Demonstration of the radiative and convective zones in atmospheres of irradiated vs non-irradiated giant planets. A Jupiter-sized planet is considered with solar elemental abundances in chemical and radiative-convective equilibrium. The irradiated planet has an equilibrium temperature of $1000$ K (left) and the one with no irradiation has an internal flux corresponding to $1000$ K (right). The solid lines are the converged pressure-temperature profiles, the blue shaded region represents the radiative zone, and the red region near the bottom of the atmosphere is the convective zone, with the transition point shown by the black circle.}
  \label{fig:convection_example}
\end{figure}

\subsubsection{Convection}
\label{sec:convection}
It may be the case that some regions of the atmosphere can be unstable to convection. Convection needs to be considered since this will almost always dominate the energy transport mechanism over radiative transport deeper in the atmosphere. Here the optical depth can be high and the radiative flux can be low and inefficient to transport energy. In order to account for convection, the region  where the atmosphere is convective needs to be determined and the appropriate flux needs to be calculated. 

Convection is treated in the model using the Mixing Length Theory \citep{kippenhahnbook}. We can assume adiabatic cooling for a parcel of gas that rises up in the atmosphere (the time scale for the gas to be in thermal equilibrium with the surroundings is negligible compared to the time it takes to rise in the atmosphere for most cases). Considering gas giant atmospheres consist mainly of diatomic hydrogen and the temperature is usually such that the 3 rotational modes are active, so $\gamma \equiv C_p/C_V=7/5$. Here we assume that the vibrational modes are suppressed, however, this is easy to modify if required. The adiabat is $PV^\gamma=\mathrm{cst}$, so we have $(d \mathrm{ln}T/d \mathrm{ln}P)|_{\mathrm{ad}} = 2/7$ from the ideal gas equation. If the temperature gradient exceeds the adiabatic gradient (or dry lapse rate) then the parcel of gas will be warmer than the surroundings it will rise. The temperature gradient thus determines which regions of the atmosphere, if any, will be unstable to convection. If the Schwarzschild condition for convective instability holds,
\begin{align}
\nabla > \nabla_\mathrm{ad},
\end{align}
the radiative equilibrium conditions \ref{eqn:reqm_int_no_conv} and \ref{eqn:reqm_diff_no_conv} now include another flux where the atmosphere is unstable to convection.
\begin{align}\label{eqn:reqmint}
\int_0^\infty \kappa_\nu (J_\nu-B_\nu) d\nu + \frac{\rho g}{4\pi}\frac{dF_\mathrm{conv}}{dP} & = 0,\\ 
\int_0^\infty \frac{d(f_\nu J_\nu)}{d\tau_\nu} d\nu +\frac{F_\mathrm{conv}}{4\pi} & = \frac{\sigma_R}{4 \pi}T_\mathrm{int}^4,
\label{eqn:reqmdiff}
\end{align}
with the convective flux $F_\mathrm{conv}$ given by mixing length theory (full details of the derivations can be found in \citealt{kippenhahnbook}).
\begin{align}
F_{\mathrm{conv}}(\nabla,T,P) & = \left(\frac{gQH_P}{32}\right)^{1/2}\rho c_P T (\nabla - \nabla_{\mathrm{el}})^{3/2} (l/H_P)^2, \\
& \equiv F_0(\nabla - \nabla_{\mathrm{el}})^{3/2}, \nonumber \\
\nabla-\nabla_{\mathrm{el}} & = \frac{1}{2}B^2 + (\nabla-\nabla_{\mathrm{ad}}) -B\left(\frac{1}{4}B^2 + (\nabla-\nabla_{\mathrm{ad}})\right)^{1/2}, \\
B & \equiv \frac{16\sqrt{2}\sigma_R T^3}{\rho c_P (gQH_P)^{1/2}(l/H_P)}\frac{\tau_{el}}{1+ \frac{1}{2}\tau_{el}^2},
\end{align}
where $\nabla \equiv d\mathrm{ln}T/d\mathrm{ln}P$, $c_p$ is the heat capacity at constant pressure and $Q \equiv -(d\mathrm{ln}\rho/d\mathrm{ln}T)_P$, which is equal to 1 for an ideal gas. $\tau_{\mathrm{el}} = l \chi_R$ is the optical depth of a small parcel of gas of size $l$, which is a free parameter usually taken to be $\approx H_P$, the scale height of the atmosphere. The choice of $l$ has  minimal effect on the observed flux for hot Jupiters as convective regions are generally deep below the observable atmosphere as discussed in section~\ref{sec:results}. $\chi_R$ is the Rosseland mean opacity and $\nabla_{\mathrm{el}}$ is the elemental logarithmic temperature gradient, which satisfies 
\begin{align}
\nabla_{\mathrm{el}}-\nabla_{\mathrm{ad}} &= B ~\sqrt[]{\nabla-\nabla_{\mathrm{el}}}. 
\end{align}
By adding $\nabla$ to both sides and rearranging, $\nabla-\nabla_{\mathrm{el}}$ can be calculated from the resultant quadratic in $\sqrt[]{\nabla-\nabla_{\mathrm{el}}}$.
\begin{align}
\nabla-\nabla_{\mathrm{ad}} &= \nabla-\nabla_{\mathrm{el}} + B ~\sqrt[]{\nabla-\nabla_{\mathrm{el}}}.\label{eqn:grad_el_grad_ad}
\end{align}

The computational method to implement this calculation is described in the appendix. If convection is present in the atmosphere, the convective flux terms are also linearised by Rybicki's method described below.


\subsubsection{Rybicki's Method and Linearisation}\label{Rybicki_method}

We perturb the temperature profile iteratively to converge to radiative-convective equilibrium using Rybicki's Method and Linearisation. Using the Feautrier method described in section~\ref{feautrier method} above the solution to the radiative transfer equation is obtained for a given temperature. To proceed and find a correction to the P-T profile the equations above (equations \ref{eqn:rtJ}, \ref{eqn:topbc}, \ref{eqn:bottombc}, \ref{eqn:reqmint} and \ref{eqn:reqmdiff}) are linearised with respect to temperature. To implement Rybicki's correction procedure \citep{rybicki_paper}, all elements ($J$, $T$, $\Delta \tau$, $\kappa$ and $\sigma$, $B$ etc.) are replaced with perturbed values with respect to $J_k$ (where applicable for a given frequency $k$) and $T$ (e.g $J_k$ to $J_k+\delta J_k$, $T$ to $T+\delta T$, $\Delta \tau$ to $\Delta \tau + \frac{\delta \Delta \tau}{\delta T} dT$, $B$ to $B +\frac{\delta B}{\delta T}dT$ etc.). The linearised absorption coefficient is 
\begin{align}
\delta \kappa & = \frac{\delta \kappa}{\delta T}dT = \left(\frac{\partial \kappa}{\partial T} + \frac{\partial \kappa}{\partial P}\frac{dP}{dT}\right) \, dT,
\end{align}
and similarly for the scattering coefficient. The full details of the linearisation of all coefficients can be found in \cite{hubenybook}, but is essentially a Newton-Raphson procedure. It is described in more detail in the appendix.
Introducing the vector $\mathbf{\delta J_k} = (\delta J_{1,k},\delta J_{2,k}, \cdots, \delta J_{ND,k})$ one obtains the set of matrix equations for $J_k$ and $T$
\begin{align}
\textbf{U}_k\delta \textbf{J}_k + \textbf{V}_k \delta \textbf{T} & = \textbf{E}_k, \label{eqn:uve}\\
\sum_{k=1}^{NF}\textbf{X}_k\delta \textbf{J}_k + \textbf{A} \delta \textbf{T} & = \textbf{F},\label{eqn:xaf}
\end{align}
where $NF$ is the total number of frequencies. \textbf{U} and \textbf{V} are the tridiagonal matrices obtained from linearisation of RTE, and \textbf{X} and \textbf{A} are the bidiagonal matrices from linearisation of radiative equilibrium. equation \ref{eqn:uve} represents the linearised form of the radiative transfer equation for every frequency $k$, and equation \ref{eqn:xaf} is the constraint of radiative-convective equilibrium (either form). Combining each of the equations \ref{eqn:uve} (i.e., for every frequency) and equation \ref{eqn:xaf} together yields the matrix equation
\begin{align}\label{eqn:matrixeqn}
 \begin{pmatrix}
  \mathbf{U_1} & 0 & \cdots & \cdots & 0 & \mathbf{V_1} \\
  0 & \mathbf{U_2} & 0 & \cdots  & \vdots & \mathbf{V_2} \\
  \vdots & 0 & \mathbf{U_3}  & 0 & \vdots & \mathbf{V_3}\\
  \vdots  & \vdots  & \vdots & \ddots & \vdots & \vdots\\
  0 & 0 & 0 & \cdots & \mathbf{U_{NF}} & \mathbf{V_{NF}} \\
  \mathbf{X_1} & \mathbf{X_2} & \mathbf{X_3} & \cdots & \mathbf{X_{NF}} & \mathbf{A}
 \end{pmatrix}
 \begin{pmatrix}
\delta \mathbf{J_1} \\ \delta \mathbf{J_2} \\ \delta \mathbf{J_3} \\ \vdots \\ \delta \mathbf{J_{NF}} \\ \delta \mathbf{T}
 \end{pmatrix}
 & = 
 \begin{pmatrix}
\mathbf{E_1} \\ \mathbf{E_2} \\ \mathbf{E_3} \\ \vdots \\ \mathbf{E_{NF}} \\ \mathbf{F}
 \end{pmatrix}
\end{align}

The vector $\delta \mathbf{J_k}$ can be written in terms of $\delta \mathbf{T}$ by
\begin{align}
\delta \mathbf{J_k} = (\mathbf{U_k^{-1}E_k}) - (\mathbf{U_k^{-1}V_k})\delta \mathbf{T},
\end{align}
Inversion of the $\mathbf{U_k}$ matrices is only requires a linear number of operations in the number of layers $ND$, as they are tridiagonal. The bottom line of the matrix equation \ref{eqn:matrixeqn} is then given by
\begin{align}
\sum_{k=1}^{NF} \mathbf{X_k}((\mathbf{U_k^{-1}E_k}) &- (\mathbf{U_k^{-1}V_k})\delta \mathbf{T}) + \mathbf{A} \delta \mathbf{T} = \mathbf{F} , \nonumber \\
\left(\mathbf{A} - \sum_{k=1}^{NF} \mathbf{X_k(U_k^{-1}V_k)}\right) \, \delta \mathbf{T} &= \left(\mathbf{F} - \sum_{k=1}^{NF} \mathbf{X_k(U_k^{-1}E_k)}\right).\label{eqn:dTsolution}
\end{align}
The inversion of the $ND\times ND$ matrix on the left gives the change in temperature of each layer by yielding $\delta \mathbf{T}$. As the matrix on the left is in general full, this is an $O(ND^3)$ operation. Hence the total computation time $a(ND)(NF)+b(ND^3)$ is highly advantageous as it is linear in the frequencies, and one can have a very large number and sample many wavelengths without an overly expensive increase in computation time. The new temperature is then found for each layer by applying the correction $\delta \mathbf{T}$ to $\mathbf{T}$. The whole process is then repeated for this new temperature, iteratively until a tolerance of $\delta\mathbf{T}/\mathbf{T}$ = $10^{-4}$ is reached. 

\subsection{Chemical Composition}\label{eqm_chem}

The chemical composition centrally governs the opacity in each layer of the atmosphere. The opacity due to any chemical species is given by the product of its absorption cross section and its number density in the given layer. The total opacity is then the sum of contributions from all individual species. In section~\ref{opac_calcs}, we discuss how we compute the absorption cross sections for the different chemical species. Here we discuss how we compute their abundances. 

The abundances of the chemical species in the atmosphere are determined under the assumption of thermochemical equilibrium. In planetary atmospheres the chemical species are mostly molecular with the exception of a few species which can survive in atomic form under specific conditions. The equilibrium abundances of the various chemical species are determined by minimizing the Gibbs free energy of the system for a given temperature, pressure, and elemental abundances. There is an extensive body of literature on the chemical compositions of exoplanetary atmospheres under various conditions \citep{lodders2002,venot2012,Madhuco_2012,moses2013,blecic2016,Heng_chem_2016,Madhu_review2016}. In general, the equilibrium abundances of a large number of species can be determined numerically for any $P$ and $T$, considering a full list of elements. However, the chemical abundances and opacity in giant exoplanetary atmospheres are dominated by species containing O, C, and N, which are the most abundant elements after H and He. Therefore, in the present initial study, we consider equilibrium abundances of only the dominant chemical species containing O, C, and N. We denote the abundances by volume mixing ratios, i.e. ratios by number density. 

\begin{figure}
  \centering
    \includegraphics[width=\columnwidth]{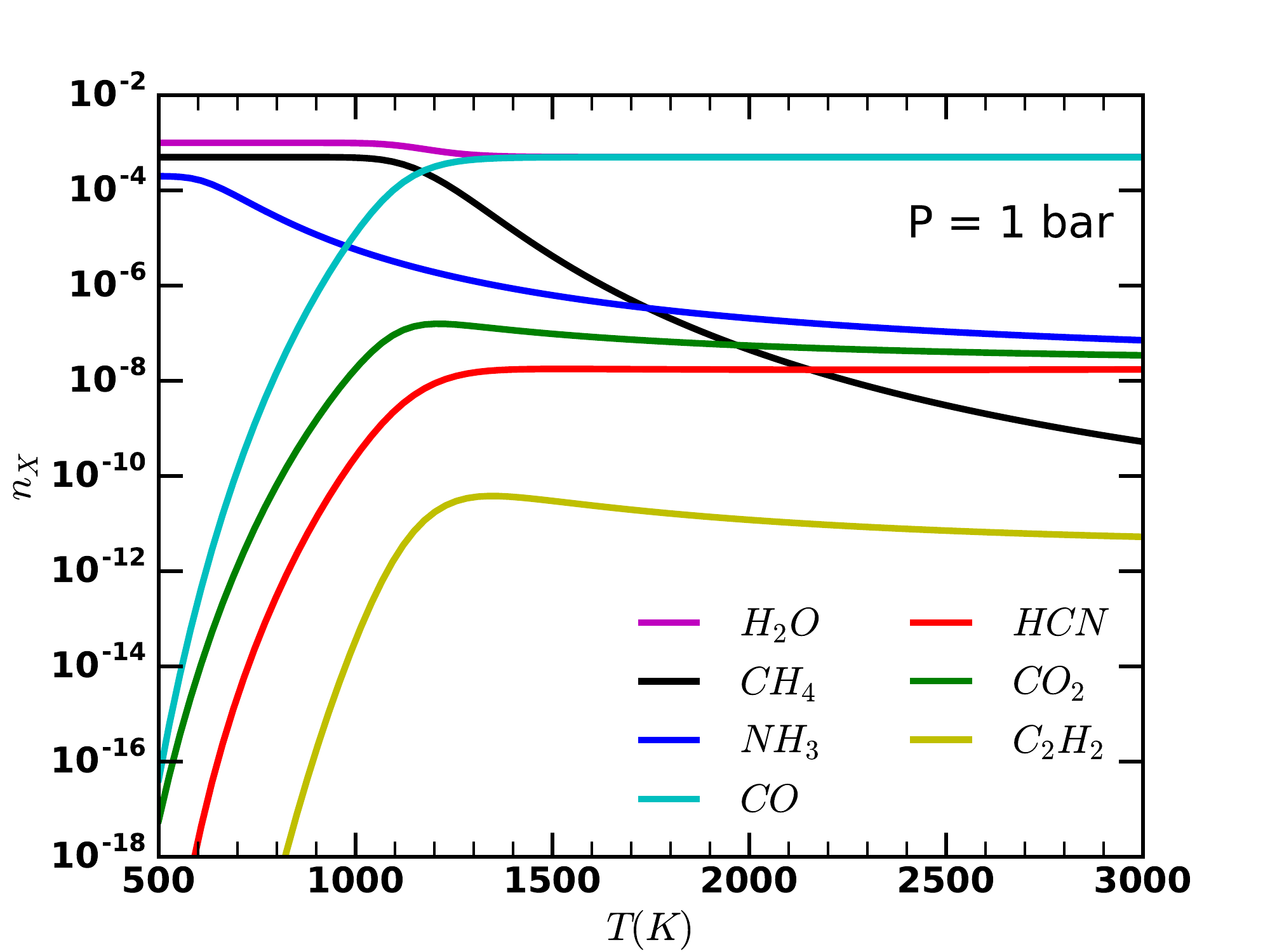}
    \caption{Molecular mixing ratios of prominent molecules in H$_2$-rich atmospheres in chemical equilibrium. The dependence of the mixing ratios on the temperature is shown for a nominal pressure of 1 bar assuming solar elemental abundances, as discussed in section~\ref{eqm_chem}.}
  \label{fig:eqm_chem}
\end{figure}

In the present work, we focus on giant planet atmospheres. Therefore, the key parameters driving the chemical mixing ratios are $P$, $T$, and the elemental abundances of He, O, C, and N, relative to H. Our model atmospheres span a wide range in metallicities (solar to 30$\times$ solar), C/O ratios (0.5-1.5), $P$ ($10^{-5}$ - 100 bar), and $T$ ($\sim$400-3000 K). While we can compute the chemical mixing ratios numerically \citep{Madhuco_2012}, we currently use the semi-analytical prescriptions of \citep{Heng_chem_2016} which give accurate estimates for the prominent O, C, and N based molecules of interest over the desired parameter space. These molecules are H$_2$O, CH$_4$, NH$_3$, CO, HCN, CO$_2$, C$_2$H$_2$, N$_2$ and C$_2$H$_4$. In specific cases, we also consider additional gaseous species such as Na, K, and TiO, to demonstrate particular physical processes, e.g. of strong optical absorption or  thermal inversions. Here we briefly discuss the prescriptions we use for the prominent molecules. Figure~\ref{fig:eqm_chem} shows a representative calculation of molecular mixing ratios for solar elemental abundances at a nominal pressure of 1 bar and over a wide range of temperatures. 

The main 6 reactions that govern the mixing ratios of the O, C, N molecules are \citep{Heng_chem_2016}:
\begin{equation}
\begin{split}
\mbox{CH}_4 + \mbox{H}_2\mbox{O} &\leftrightarrows \mbox{CO} + 3\mbox{H}_2 , \\
\mbox{CO}_2 + \mbox{H}_2 &\leftrightarrows \mbox{CO} + \mbox{H}_2\mbox{O},\\
2\mbox{CH}_4 &\leftrightarrows \mbox{C}_2\mbox{H}_2 + 3\mbox{H}_2, \\
2\mbox{C}_2\mbox{H}_4 &\leftrightarrows \mbox{C}_2\mbox{H}_2 + \mbox{H}_2, \\
2\mbox{NH}_3 &\leftrightarrows \mbox{N}_2 + 3\mbox{H}_2,\\
\mbox{NH}_3 + \mbox{CH}_4 &\leftrightarrows \mbox{HCN} + 3\mbox{H}_2.\\
\end{split}\nonumber
\end{equation}
Taking $n_i = N_i/N_{H_2}$ to be the mixing ratio of species $i$ relative to hydrogen, where $N_i$ is the number density. The particle conservation equations give 
\begin{align*}
n_{CH_4}+ n_{CO} +n_{CO_2} + n_{HCN} + 2n_{C_2H_2} +2n_{C_2H_4} &= 2n_C,\\
n_{H_2O}+ n_{CO} +2n_{CO_2} &= 2n_O,\\
2n_{N_2}+ n_{NH_3}+ n_{HCN} &= 2n_N.
\end{align*}
The equilibrium constants are
\begin{align*}
K_1 &= \frac{n_{CO}}{n_{CH_4}n_{H_2O}} = \left(\frac{P_0}{P}\right)^2\mathrm{exp}\left(\frac{-\Delta G_{0,1}}{RT}\right),\\
K_2 &= \frac{n_{CO}n_{H_2O}}{n_{CO_2}} = \mathrm{exp}\left(\frac{-\Delta G_{0,2}}{RT}\right),\\
K_3 &= \frac{n_{C_2H_2}}{n_{CH_4}^2} = \left(\frac{P_0}{P}\right)^2\mathrm{exp}\left(\frac{-\Delta G_{0,3}}{RT}\right),\\
K_4 &= \frac{n_{C_2H_2}}{n_{C_2H_4}} = \left(\frac{P_0}{P}\right)\mathrm{exp}\left(\frac{-\Delta G_{0,4}}{RT}\right),\\
K_5 &= \frac{n_{N_2}}{n_{NH_3}^2} = \left(\frac{P_0}{P}\right)^2\mathrm{exp}\left(\frac{-\Delta G_{0,5}}{RT}\right),\\
K_6 &= \frac{n_{HCN}}{n_{NH_3}n_{CH_4}} = \left(\frac{P_0}{P}\right)^2\mathrm{exp}\left(\frac{-\Delta G_{0,6}}{RT}\right).\\
\end{align*}
Here, $\Delta G_{0,i}$ is the standard Gibbs free energy for reaction $i$ given in \cite{Heng_chem_2015}. Specifying the ratios $C/H = n_C$, $O/H=n_O$ and $N/H=n_N$, the 9 equations can then be solved for the mixing ratios of the 9 species.  The details of the implementation can be found in \citet{Heng_chem_2016}. Thus, given a P-T profile the equilibrium mixing ratios of all the 9 molecules can be determined in each layer. The mixing ratios can then be multiplied by the number density of H$_2$ to obtain the total number density of gas $N_i$ of each species $i$. The number density together with the relevant cross sections can then be used to obtain the total absorption and opacity, as discussed below.  

\subsection{Opacities}\label{opac_calcs}

In this section we discuss the sources of the opacity data in our model atmosphere. The extinction in our model is due to absorption from gaseous species, collisionally induced absorption from H$_2$-H$_2$ and H$_2$-He and Rayleigh scattering from H$_2$. The chemical species considered and their abundances in equilibrium where discussed in the previous section. Here, we discuss how we compute the absorption cross sections of the species. The absorption coefficient $\kappa_{\nu,i}$ for a species $i$ is related to the absorption cross section $\alpha_{\nu,i}$ as $\kappa_{\nu,i} = N_i\alpha_{\nu,i} = n_iN_{H_2}\alpha_{\nu,i}$. The total absorption coefficient in a given layer of the atmosphere is the sum of the contributions from all the different species. The frequency-dependent absorption cross sections of a given molecule depend on both the temperature and pressure which contribute to the broadening of a transition line. 

Two approaches have been used for opacity calculations in radiative transfer codes for modelling exoplanetary atmospheres. Some models \citep{fortney_2007, molliere_code,Malik_2016,drummond_2016} use the k-distribution method which is commonly used in models of planetary and satellite atmospheres in the solar system \citep[e.g.,][]{mckay_1989}. In this approach, the molecular cross sections are reordered in wavelength bins and resampled on a coarser grid. This allows a small number of wavelength points to sample the opacity function and, hence, lower computation times. It does however make the approximation that spectral lines of different molecules are perfectly correlated with each other in the bin, accurate when the bins chosen are small enough in frequency space. Other groups \citep{seager_code,Barman_code, burrows_2008} adopt the so called ``line-by-line" approach, using the sampled opacity cross sections directly. Although common terminology in the exoplanetary atmospheres literature, strictly speaking this is not line-by-line as the cross sections are not computed from a native line list, i.e. from all individual lines. In the stellar atmospheres literature this technique is known as opacity sampling. The details of each method and the advantages and disadvantages are discussed in \citet{hubenybook}. Although here we have used opacity sampling, GENESIS has the capability to handle arbitrary resolution down to the native resolution of the line list, as the opacity tables are simply computed to the desired resolution. For the current purpose we have chosen a resolution that is appropriate for our applications, as shown in Fig. \ref{fig:res_tests} and discussed in section \ref{sec:validation}. Any higher resolution beyond R\textasciitilde $10^4$ has a negligible effect on the spectrum and the P-T profile.

\begin{figure}
  \centering
    \includegraphics[width=\columnwidth]{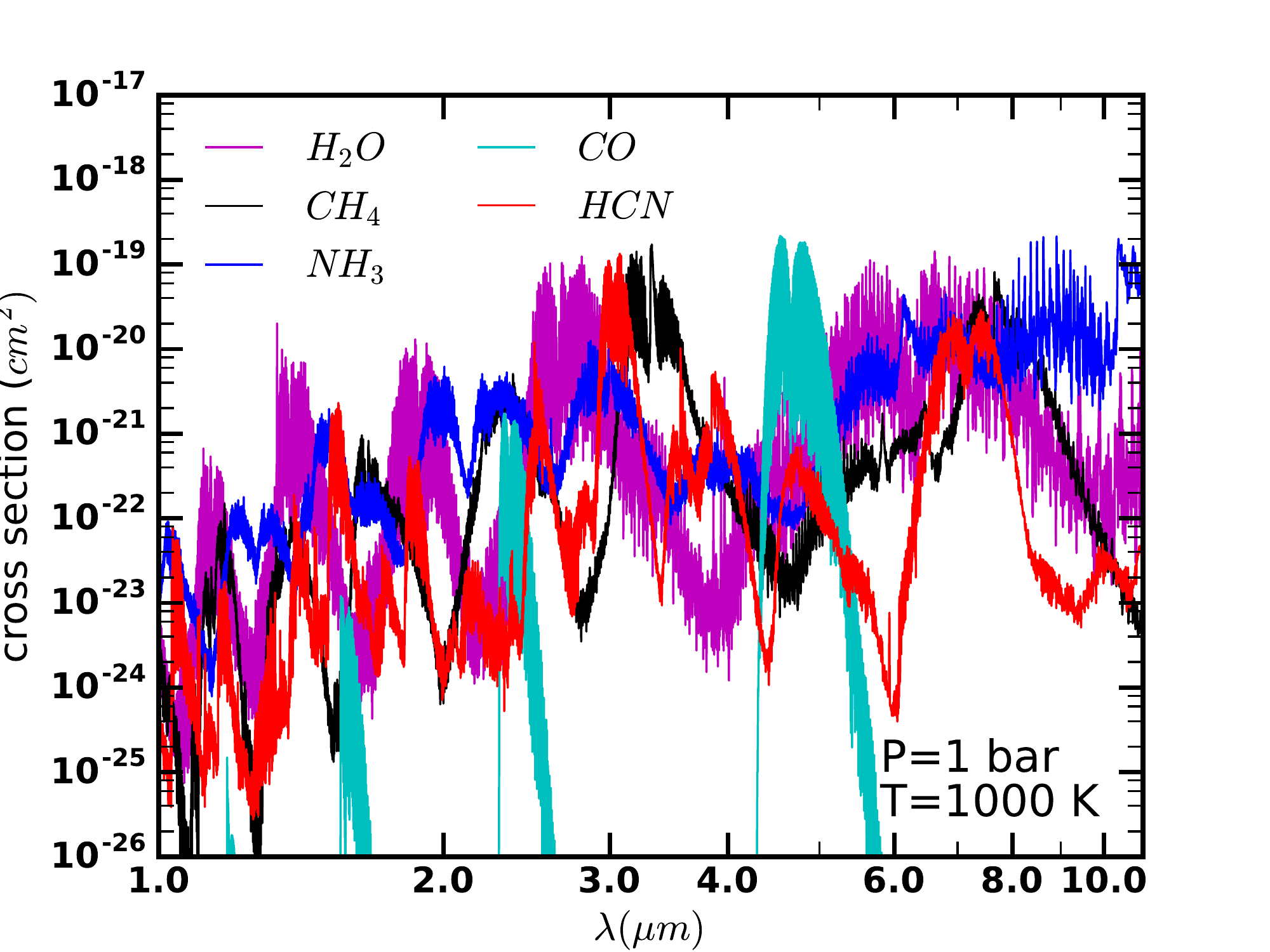}
  \caption{Molecular absorption cross sections for prominent molecules in giant planetary atmospheres considered in this work. Cross sections are shown for representative hot Jupiter conditions of T = 1000 K and P = 1 bar.}
  \label{fig:cross_sec_compare}
\end{figure}

To derive the cross sections for a molecule at a given pressure and temperature, knowledge of the line lists of transitions that can occur and the line strengths is required. In the present work we obtain the line lists from the HITEMP database for H$_2$O, CO and CO$_2$ \citep{rothman_2010} and Exomol for HCN, CH$_4$ and NH$_3$ \citep{tennyson_2016}, with the rest derived from the HITRAN line lists \citep{rothman_2013}. These databases provide high temperature line lists and their corresponding partition functions. We calculate the cross sections from these line lists following a similar method to \cite{Hedges_2016}. 

The line strengths, $S_0 \equiv S(T_\mathrm{ref})$ in the database are given at a temperature of 296K, and are converted to a general temperature using \citep{rothman_1998},
\begin{align}
S(T) &= S_0\frac{Q(T_\mathrm{ref})}{Q(T)}\frac{\mathrm{exp}(-E_\mathrm{lower}/k_bT)}{\mathrm{exp}(-E_\mathrm{lower}/k_bT_{\mathrm{ref}})}\frac{1-\mathrm{exp}(-h\nu_0/k_bT)}{1-\mathrm{exp}(-h\nu_0/k_bT_{\mathrm{ref}})},
\end{align}
where $E_\mathrm{lower}$ is the lower energy state of the transition, $\nu_0$ is the frequency of the transition line ($E_\mathrm{upper}-E_\mathrm{lower} = h\nu_0$) and $Q$ is the partition function,
\begin{align}
Q(T) &= \sum_j g_j \mathrm{exp}(-E_j/k_bT)
\end{align}
with the degeneracy of the state $j$ given by $g_j$ and where $k_b$ is the Boltzmann constant.

Once the line strength is obtained, the broadening of the line needs to be taken into account. At a given temperature, thermal motion of the molecule from a Maxwell-Boltzmann type distribution will result in a Doppler shift of the line. This will blur out the transition line, so instead of a sharp absorption peak, a Gaussian shape will be observed. For a molecule of mass $m$, the Gaussian profile $f_D$ is given by
\begin{align}
f_D(\nu-\nu_0) &= \frac{1}{\gamma_G \sqrt[]{\pi}}\mathrm{exp}\left(-\frac{(\nu-\nu_0)^2}{\gamma_G^2}\right),\\
\gamma_G &\equiv \sqrt[]{\frac{2k_bT}{m}}\frac{\nu_0}{c}.
\end{align}
In this case the full width half maximum of the gaussian is given by $\sqrt[]{2}\gamma_G$, and the frequency $\nu$ and the central line frequency $\nu_0$ are given in $\mathrm{cm}^{-1}$. The line will also become broadened due to the pressure of the gas. This will result in a Lorentzian profile,
\begin{align}
f_P(\nu-\nu_0) &= \frac{1}{\pi}\frac{\gamma_L}{(\nu-\nu_0)^2+\gamma_L^2},\\
\gamma_L &\equiv \left(\frac{T_{\mathrm{ref}}}{T}\right)^n P \sum_b \gamma_{L,b}\, p_b,
\end{align}
with $\gamma_{L,b}$ the Lorentzian HWHM from a specific broadening molecule, and $n$ a temperature scaling factor. Pressure broadening can be more difficult to calculate, as the parameters $n$ and $\gamma_{L,b}$ are needed, and depend not just on the line, but also the main constituent of the atmosphere. Given that we are modelling giant planet atmospheres, ideally we would need pressure broadening parameters due to H$_2$. However, such data is only now becoming available and for only a couple of molecules \citep{wilzewski2015,barton2016}. Therefore, in the present work we have instead used parameters for air broadening for all the molecules for a uniform analysis. We tested for H$_2$O cross sections with H$_2$ broadening to find that the differences are not significant for the current analysis. Nevertheless, our opacity database is continually updated with new broadening data. 

\begin{figure*}
 \includegraphics[trim = 40mm 0mm 45mm 0mm, clip,width=\textwidth]{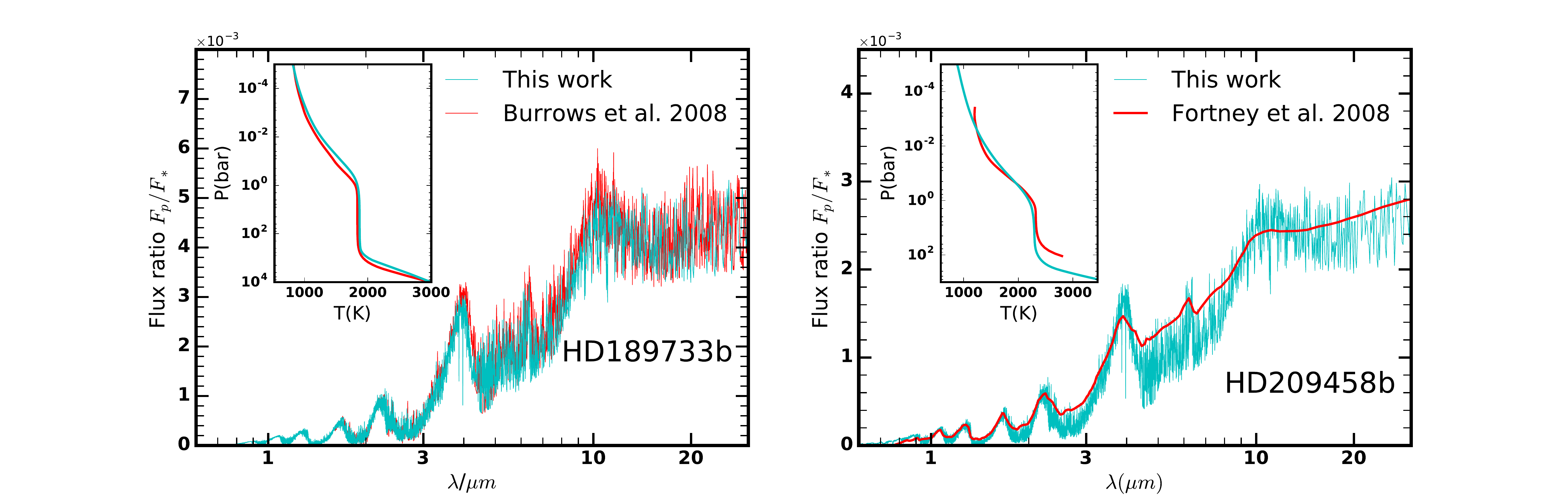}
 \caption{Comparison of GENESIS models with published results. A model of the hot Jupiter HD~189733~b from Fig. 1 and Fig. 4 of \citealt{burrows_2008} is shown in the left panel, and a model of HD~209458~b from Fig. 12 of \citealt{fortney_2008} is shown on the right. The planet-star flux ratios and the P-T profiles in the inset show good agreement. The models assume chemical equilibrium with solar elemental abundances.}
\label{fig:burrows_fortney}
\end{figure*}

The full broadening of a line will be the convolution of the Gaussian and Lorentzian profiles, known as the Voigt function,
\begin{align}
f_V(\nu-\nu_0) &= \int_{-\infty}^{\infty}f_D(\nu'-\nu_0)f_P(\nu-\nu')d\nu'.
\end{align}
The characteristic width of the function is given approximately by $\gamma_V \approx 0.5436\gamma_L +\sqrt[]{0.2166\gamma_L^2+\gamma_G^2}$.
Defining new variables 
\begin{align}
u&=\frac{\nu-\nu_0}{\gamma_G},\\
a&=\frac{\gamma_L}{\gamma_G},
\end{align}
the Voigt function can be cast in terms of the normalised Faddeeva function $w(z)$ \citep[see][]{grimm_2015,Hedges_2016},
\begin{align}
f_V(\nu,\gamma_L,\gamma_G) &= \frac{\mathrm{Re}(w(u+ia))}{\gamma_G \sqrt[]{\pi}}.
\end{align}
The cross section at a certain frequency $\nu$, for a particular line including broadening is then
\begin{align}
\alpha_\nu = S(T)f_V(\nu,\gamma_L,\gamma_G).
\end{align}

The $\alpha_\nu$ values were calculated on a fine resolution grid with a frequency spacing determined from the characteristic width of a line at $500 \mathrm{cm}^{-1}$ at each pressure and temperature that we considered, i.e., taking the spacing of the frequency grid to be $\gamma_V(500\mathrm{cm}^{-1},T,P)/6$, which typically is much less than 1$\mathrm{cm}^{-1}$. When this spacing proved to be larger than 1$\mathrm{cm}^{-1}$, e.g. for strong pressure broadening, a minimum spacing of 1$\mathrm{cm}^{-1}$ was used. The cut-off for the line wings was chosen to be at $\pm 250 \gamma_V$ for $P < 1$ bar and $\pm 500 \gamma_V$ for $P \geq 1$ bar for every line in the database. The reasoning behind these choices are given in detail in \cite{Hedges_2016}. 

Our cross section database spans a wide range in $P$, $T$, and frequency. Table~\ref{table:cross_sec_grid} shows the $P$ and $T$ grid over which the cross sections were computed for each molecular species, and spans the ranges relevant for giant exoplanetary and brown dwarf atmospheres. The pressure grid ranged from $10^{-4}$ bar to $100$ bar, in 6 steps equally spaced in log pressure. The temperature grid spanned 300 - 3500 K. Some of the partition functions do not go up as high as the highest temperatures considered here and in such cases an extrapolation is done by a cubic. The line-by-line cross sections were ultimately binned down and stored at a resolution of $1\mathrm{cm}^{-1}$ in the spectral range of 25000$\mathrm{cm}^{-1}$ (0.4$\micron$) to 200$\mathrm{cm}^{-1}$ (50 $\micron$). While this resolution is adequate for our current purpose higher resolution can be easily achieved as need be.   

\begin{table}
\begin{center}
\begin{tabular}{ |c| c c c c c c c |} 
\hline
 T(K) & 300&400&500 & 600&700&800&900\\
 &1000&1200&1400&1600&1800&2000&2500\\
 &3000&3500&&&&& \\ 
 \hline
 P(bar) & $10^{-4}$ & $10^{-3}$ & $10^{-2}$ & $10^{-1}$&1&10&100 \\ 
 \hline
\end{tabular}\\
\caption{The temperature and pressure grid for the cross sections, which are the same as those used in \citet{Hedges_2016}.}\label{table:cross_sec_grid}
\end{center}
\end{table}

As well as molecular absorption, collisionally induced absorption from the H$_2$ and He rich atmosphere is also required. This H$_2$-H$_2$ and H$_2$-He absorption is taken from the HITRAN database and included with the mixing fraction of helium set as a tunable parameter. These cross sections are in a different format to the molecular cross sections, as they only vary with temperature and require multiplication by the number density squared to find the resultant opacity. At each $P$, $T$, and frequency point of our model, the relevant molecular cross sections $\alpha_{\nu,i}$ are computed for each molecule and together provide the opacity to be accounted for in the radiative transfer. In the specific cases where consider models with visible opacity from Na, K, we obtained the corresponding cross sections from \citet{burrows_sodium}. In other cases where we explore the effect of TiO in the atmospheres, we derived the cross sections from \citet{kurucz_tio}.

\subsection{Stellar Flux}

The incoming stellar flux sets the boundary condition at the top of the atmosphere (equation \ref{eqn:topbc}). We use a Kurucz model spectrum \citep{Kurucz_1979_paper,kurucz_model} based on the stellar parameters. We linearly interpolate in the effective temperature and log gravity of the Kurucz model grid to determine the theoretical stellar spectrum for the required stellar parameters at the nearest metallicity on the grid. However, with regards to the convergence of the GENESIS models and the observed flux ratio, only minor differences were seen between the stellar model and a Planck function, similar to the findings of \citet{Malik_2016} with Kurucz and PHOENIX stellar models. The incident flux on the planetary dayside for the different assumptions for stellar flux is given by
\begin{align}
F_\mathrm{ext,Planck} &= f_{\mathrm{r}} ~\pi B(T_\mathrm{eff},\nu) ~\frac{R_\mathrm{star}^2}{a^2}, \\
F_\mathrm{ext,Kurucz} &= f_{\mathrm{r}} ~4\pi H_\mathrm{star}(T_\mathrm{eff},\nu,\mathrm{log}(g_\mathrm{star}),Z) ~\frac{R_\mathrm{star}^2}{a^2}
\end{align}
where $T_{\rm eff}$ is the stellar effective temperature, $B$ is the Planck function, and $H_{\rm star}$ is the Eddington Flux at the stellar surface obtained from a Kurucz model. $R_{\mathrm{star}}$ and $a$ are the stellar radius and semi-major axis,  respectively. Here, $f_\mathrm{r}$ is used to account for the average flux incident on the whole day side of the planet, rather than only at the sub-stellar point, and to consider part of the incident flux transported to the night side. 

\begin{figure*} \includegraphics[trim = 10mm 0mm 20mm 10mm, clip,width=\textwidth]{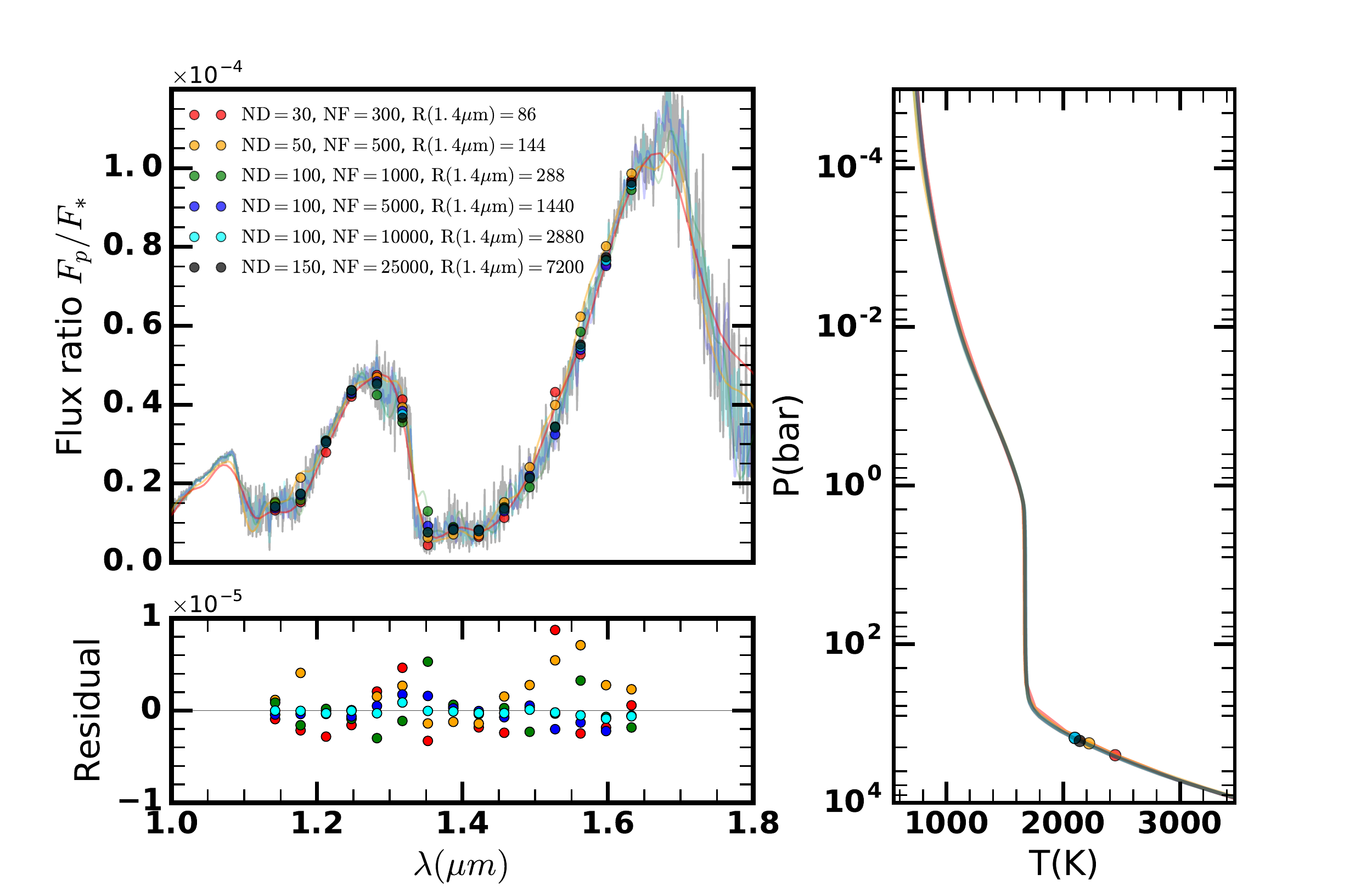}
 \caption{Model spectra and P-T profiles for a hot Jupiter with $T_{\mathrm{eq}}=1500$ K computed at different spectral and spatial resolutions. ND denotes the number of layers in the model atmosphere and NF denotes the number of frequency points with the corresponding spectral resolution $\mathrm{R} = \nu/d\nu$ given at $1.4\micron$, near the centre of a strong H$_2$O band as well as of the HST WFC3 G141 spectrograph. Six models are considered with different ND and NF, as discussed in section~\ref{sec:validation}. The left hand side (top) shows the flux ratio overlaid with circles indicating binned model points in the HST WFC3 bandpass. The bottom figure on the left shows the differences in the binned points for each model relative to the highest resolution model. The right hand plot shows the corresponding P-T profiles, with the radiative-convective boundary for each model marked with a circle of the corresponding colour.} 
\label{fig:res_tests}
\end{figure*}

\begin{figure*}
 \includegraphics[trim = 11mm 0mm 4mm 0mm, clip,width=\textwidth]{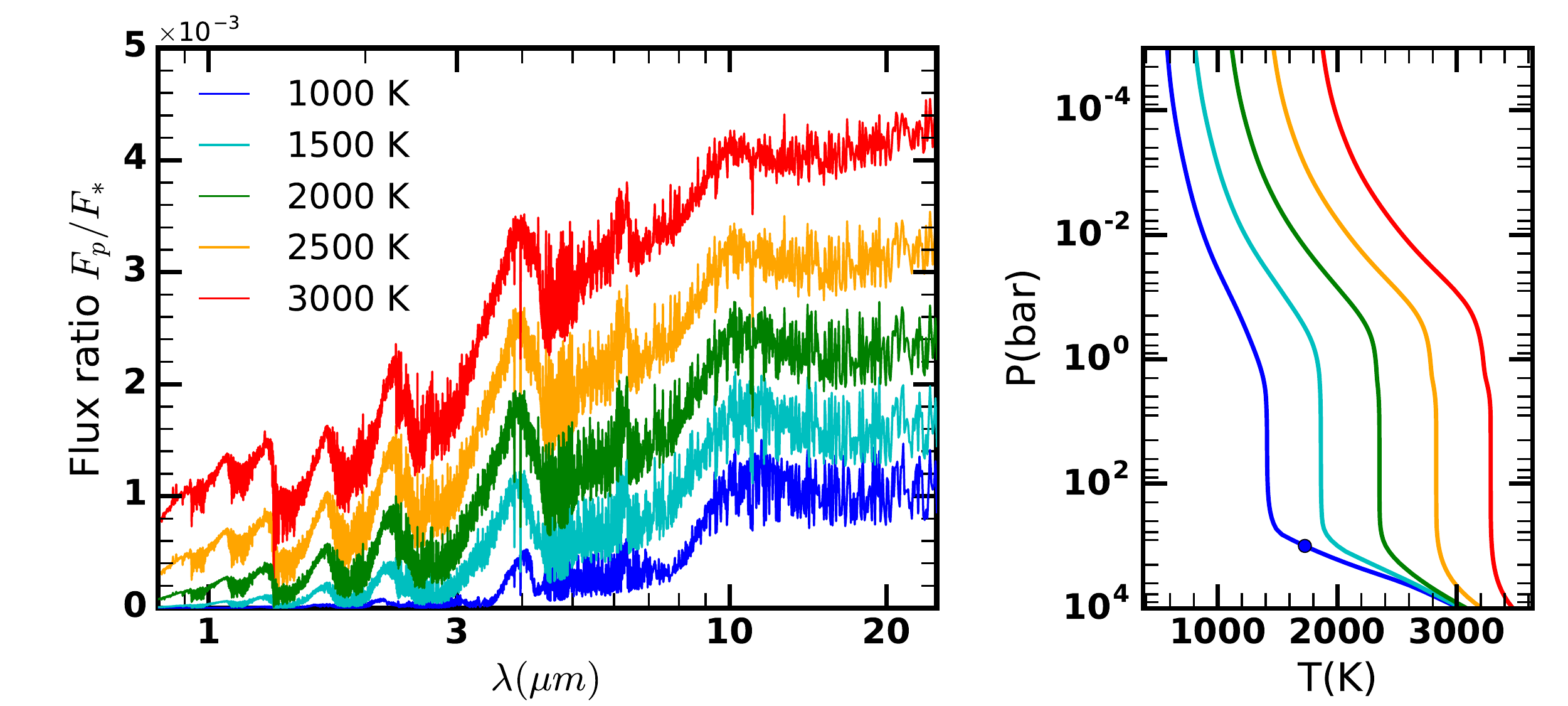}
 \caption{Self-consistent model spectra and P-T profiles of hot Jupiters with different levels of irradiation, represented by the equilibrium temperatures. The left plot shows the planet-star flux ratio as a function of wavelength for equilibrium temperatures of $T_{\mathrm{eq}}=1000K$ (blue) to $T_{\mathrm{eq}}=3000K$ (red). The right panel shows the converged P-T profiles. The mass and radius of this test planet are taken to be that of Jupiter, and the host star and the overall atmospheric metallicity is modelled as solar. The onset of convection is represented by a circle, and is only visible in the figure for the lowest equilibrium temperature; higher temperatures suppress convective regions to higher pressures due to the incident irradiation.}
 \label{fig:profiles}
\end{figure*}

\subsection{Model Validation}
\label{sec:validation}

We tested GENESIS extensively for accuracy, convergence, and performance. As discussed in section~\ref{sec:methods}, the model comprises of four key modules: (a) line-by-line opacities, (b) equilibrium chemistry, (c) radiative transfer, and (b) temperature  correction via radiative-convective equilibrium. Each of these components were tested extensively. Our new line-by-line cross sections for all the molecules were tested against the latest published sources, both our  own previous computations \citep{Hedges_2016} as well as other databases \citep{tennyson2016}, over a wide range of pressures, temperatures and wavelengths, and thermal/pressure broadening. The molecular abundances in chemical equilibrium were validated against those of \citet{Heng_chem_2016}, as discussed in section~\ref{eqm_chem}. The radiative transfer module computes a spectrum for a given P-T profile and composition. This module was extensively tested for various conditions and validated against models from our existing radiative transfer codes. Finally, the radiative-convective equilibrium module which is responsible for the iterative temperature correction procedure was tested extensively for convergence and stability; our current code is convergent to temperature differentials within our desired tolerance level of $10^{-4}$. The model is extremely stable, both in the radiative and convective regimes, over the entire range of parameter space considered in the present work. 

The GENESIS models were also tested for a range of spatial and spectral resolutions. We explored the effect of varying the number of layers in the atmosphere, $ND$, and the number of frequency points, $NF$, and hence the spectral resolution. We perform these tests in order to be able to determine what the most optimal values are for accuracy and computation time. To perform these tests, a representative hot Jupiter with an equilibrium temperature of $1500$ K was chosen around a solar type star. Fig.~\ref{fig:res_tests} reveals some interesting behaviour. Firstly, the choice of $ND$ between 30 and 150 layers has only a marginal effect on the P-T profile, with only the starting layer of the convective zone having any significant effect. This can however be explained by the discrete nature of the grid used; the temperature gradient is fixed between layers, so convective zones can only begin at layer boundaries. This is why the three test cases with $ND = 100$ have the same convective zone transition, but altering $ND$ moves the transition. The top left panel of Fig.~\ref{fig:res_tests} shows the planet-star flux ratio and binned model points in the HST WFC3 bandpass (as circles) and the bottom panel shows the differences relative to the highest resolution model. An increase in the number of frequency points used alters the binned points quite significantly until $NF=5,000$ (R$=1440$ at $1.4\micron$), where the binned data begin to lie very close to each other (\textasciitilde2 ppm); the $NF=10,000$ (R$=2880$) and $NF=25,000$ (R$=7200$) models have data points almost completely indistinguishable. Hence in the next section, all of our models were run with $NF=10,000$ and $ND=100$, as any higher resolution would be longer to run computationally but would not alter the converged P-T and flux profile significantly, as the run time scales as \textasciitilde $(NF)(ND) + \mathrm{const.}(ND^3)$.

We have also compared our full model against some published self-consistent models and found good agreement. For this purpose, we considered models of two well known hot Jupiters HD~189733b and HD~209458b. We compared our model of HD189733b with a previous cloud free model by \cite{burrows_2008}, shown in Fig.~\ref{fig:burrows_fortney}. The planetary parameters were adopted from \cite{burrows_2008} to enable an accurate comparison. We computed the molecular abundances assuming solar abundances and chemical equilibrium, and the Na and K abundances in our model increase linearly with the pressure until they reach solar values at 100 bar, to simulate their depletion at low T and P in the upper atmosphere in chemical equilibrium. The resulting converged P-T profile and planet-star flux ratio spectrum are shown in blue in Fig.~\ref{fig:burrows_fortney}, which provide a good match to those of \citet{burrows_2008} (Fig. 1 and 4 in their paper). In both cases, the convective zone starts \textasciitilde 100 bar, and the isotherm temperatures agree to within $\sim$ 15K. The P-T profile and flux are most strongly influenced by water and carbon monoxide, the only gases present in any significant quantities at these pressures and temperatures. Any slight differences are likely due to the opacities in our model being from a newer line list, or our equilibrium mixing fractions differing because of the different methods used to calculate them.

We also compared our cloud free model of HD209458b with the dayside profile from \citealt{fortney_2008} without a thermal inversion in Fig. \ref{fig:burrows_fortney}. The planetary parameters used were once again taken from \citealt{fortney_2008}. The P-T profile is matched quite well in the region where their profile was available. The model spectra also match quite well, though our model is at a higher resolution than theirs and computed with the latest opacities which are likely more complete compared to what they used. Additionally, there are significant differences in the model implementation. Our model involves line-by-line radiative transfer with the Feautrier method whereas theirs involves using the two-stream source function with the correlated-k approximation, implemented at a lower resolution which gives the smoother spectrum seen in Fig.~\ref{fig:burrows_fortney}. Similarly the temperature correction scheme is also different as discussed in section~\ref{sec:introduction}. Nevertheless, the good agreement between the two models both in the P-T profile and the spectrum is noteworthy, particularly given the development of more complete line lists for molecules over the recent years (see section \ref{opac_calcs}). 

\section{Results}
\label{sec:results}
We now use the methods discussed above to generate models of exoplanetary atmospheres over a range in parameter space. 
We assume some nominal values for the system parameters for our baseline model and explore variations thereof for our model grid. The model input parameters are shown in Table~\ref{tab:params}. Firstly, except for our models of known exoplanets, we  generally assume the planetary mass and radius, and hence gravity, to be Jovian-like and the stellar properties to be solar.  Secondly, while our model is generally applicable to any composition, we explore only H$_2$-dominated atmospheres, i.e., such as atmospheres of gas giants and ice giants. Thus, our baseline model has a solar elemental composition and we explore variations thereof in metallicities and C/O ratios for other models. The key sources of opacity are the prominent molecules expected in thermochemical equilibrium for the given elemental abundances, collision-induced absorption from H$_2$-$H_2$ and H$_2$-He, and Rayleigh scattering from H$_2$. For the incident stellar flux, in the cases of planets with strong irradiation, we assume that half the incident flux is redistributed to the night side, i.e. $f_{\mathrm{r}}$ of 0.5. For the isolated planets we assume an internal heat flux corresponding to a blackbody with $T_\mathrm{int} = 1500$ K, and for the irradiated planets we assume $T_\mathrm{int}=75$ K. 

For each GENESIS model, we report the emergent spectrum and the atmospheric pressure-temperature (P-T) profile as a function of a chosen parameter. All the models are computed over a wavelength range of 0.4-30$\micron$ wavelength, with $NF=10,000$ evenly spaced points in frequency, and $ND=100$ layers in the atmosphere evenly spaced in log P. We explore a wide range of atmospheric models in two regimes of importance for currently known exoplanets: (a) highly irradiated close-in exoplanets which dominate the transiting planet population, and (b) planets with negligible irradiation dominated by internal flux, as relevant for directly imaged exoplanets on wide orbital separations. We also generate model spectra and temperature profiles of several irradiated hot Jupiters over a wide temperature range, and compare our models of two well studied hot Jupiters to some of those reported in the literature.

\begin{figure*}
 \includegraphics[trim = 36mm 0mm 42mm 0mm, clip,width=\textwidth]{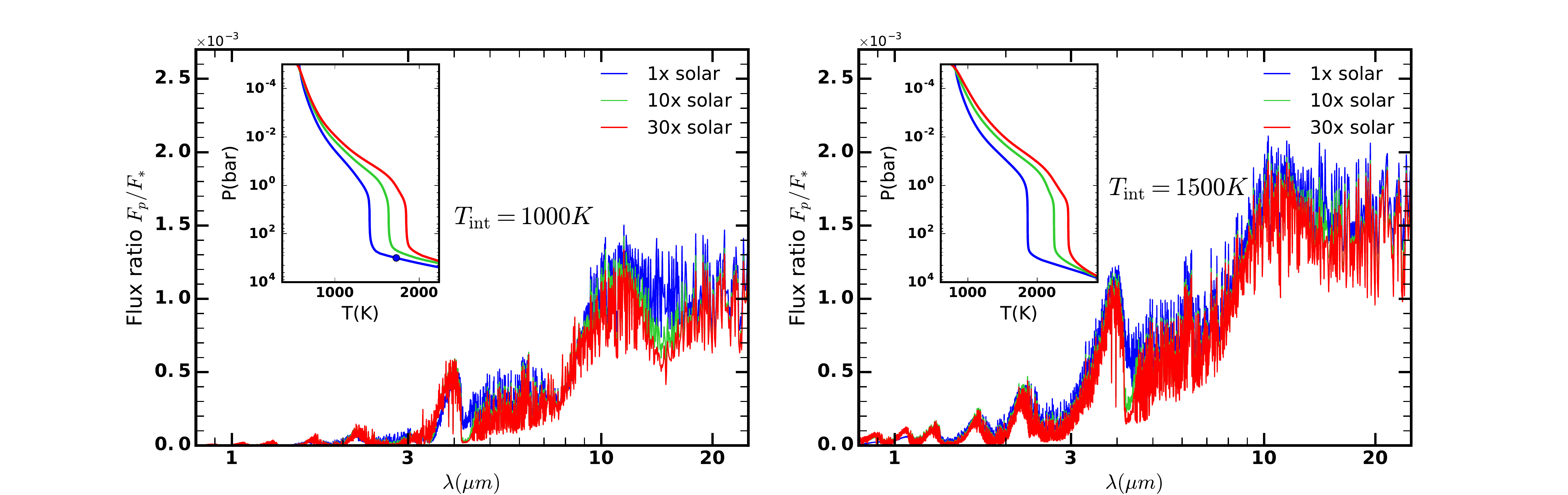}
 \caption{Effect of metallicity on the planet-star flux ratio and P-T profile of a hot Jupiter. The metallicity is explored between 1-30$\times$ solar, with the left panel showing models with equilibrium temperature of $1000$K and the right showing $1500$K. The inset plots show the converged P-T profiles.}
\label{fig:metallicity}
\end{figure*}

\subsection{Irradiated Atmospheres}
We first investigate atmospheric models of highly irradiated giant planets, particularly hot Jupiters, which are the most studied class of exoplanetary atmospheres. Unless specified otherwise we assume a Jovian-like planet, in mass and size, orbiting a sun-like star. 
We explore the dependance of emergent spectra and temperature profiles on the degree of irradiation, metallicity and C/O ratio, and the presence of optical absorbers which dominate the absorption of incident starlight and can cause thermal inversions. For the spectra, we report the planet-star flux ratio as a function of wavelength as typically measured for transiting hot Jupiters \citep[see e.g.][]{Madhu_review2014}. We assume a Kurucz model spectrum for the star. 

\begin{figure*}
 \includegraphics[trim = 36mm 0mm 42mm 0mm, clip,width=\textwidth]{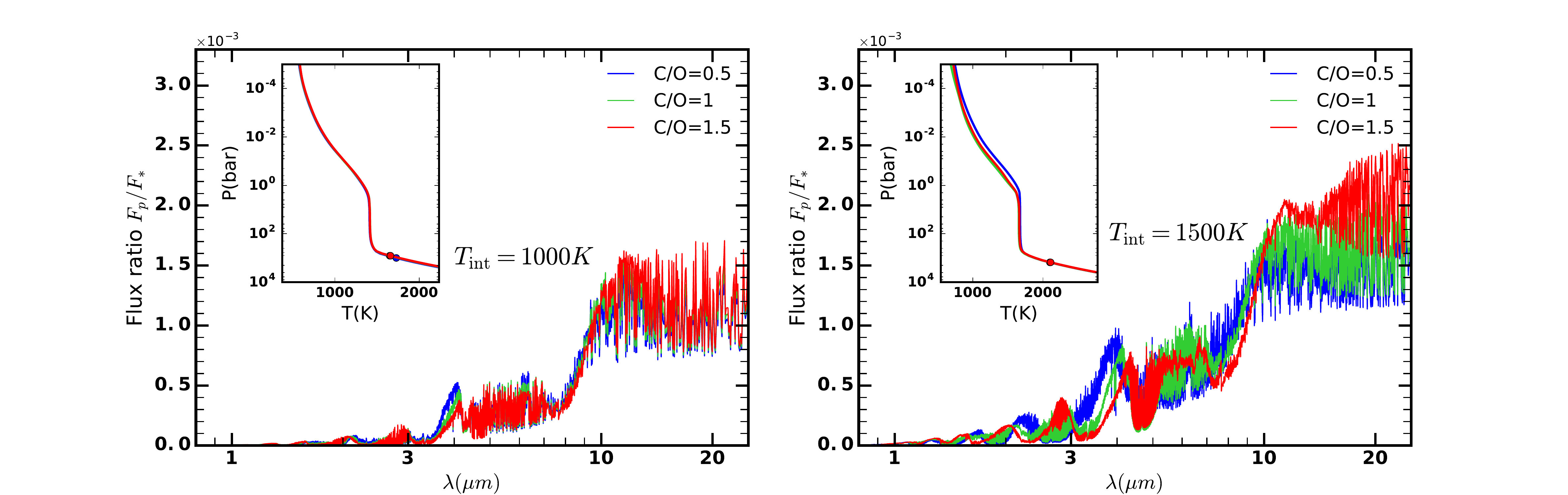}
 \caption{Effect of C/O ratio on the planet-star flux ratio and P-T profile of a hot Jupiter. Two models are considered, with equilibrium temperatures of $1000K$ (left) and $1500K$ (right). The O/H is fixed at solar value and the C/H is varied to obtain the required C/O ratio; the solar C/O ratio is 0.5.}
\label{fig:co_ratio}
\end{figure*}

\subsubsection{Effect of Stellar Irradiation}
\label{sec:irradiation}
We begin with the test case of a hot Jupiter around a sun-like star and investigate models with varying degrees of irradiation, represented by the equilibrium temperature $T_{\mathrm{eq}}$. Altering the semi-major axis, and hence the incident irradiation, causes the equilibrium temperature to vary according to $T_{\mathrm{eq}} = T_{\mathrm{eff}}~\sqrt[]{R_{\rm star}/2a}$, assuming efficient energy redistribution between the day and night sides. We consider models with $T_\mathrm{eq}$ between 1000 K and 3000 K (in steps of 500 K), representing the temperature range of currently known hot Jupiters. The chemical composition is fixed by thermochemical equilibrium assuming solar elemental abundances and considering only prominent C-N-O chemistry, as discussed in section~\ref{eqm_chem}. Depending on the particular temperature, the key sources of opacity are the prominent O and C bearing species such as H$_2$O, CO, CH$_4$, and CO$_2$ which contribute significant molecular absorption primarily in the infrared. We also included Na and K as sources of visible opacity with abundances as discussed in section~\ref{sec:validation}.

\begin{figure*}
 \includegraphics[trim = 36mm 0mm 42mm 0mm, clip,width=\textwidth]{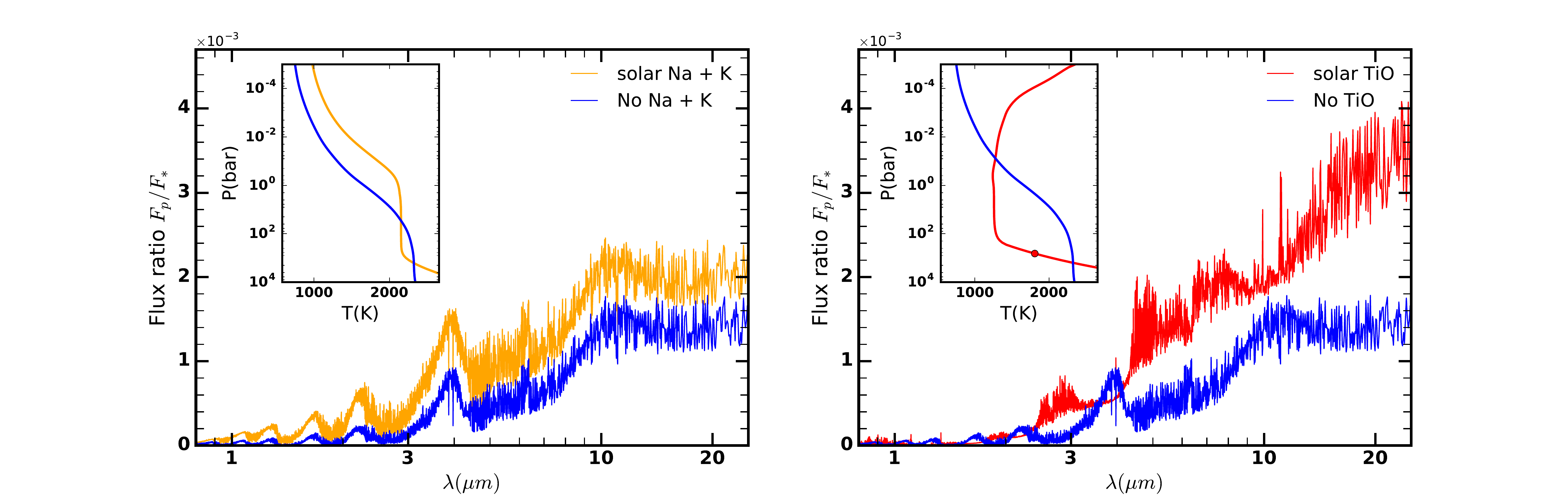}
 \caption{{Effect of Na/K and TiO visible opacity on the emergent spectrum and P-T profile of a hot Jupiter.} The left plot shows a hot Jupiter without sodium or potassium (blue) and with sodium+potassium included at solar abundances (yellow). The right hand side shows the same blue curve (without any visible absorbers) and the red curve with TiO at solar abundances. In each case the equilibrium temperature was $1500K$, with an identical incident stellar flux for all the cases shown. The inset  shows the converged radiative-convective equilibrium P-T profiles. The transition to the convective zone for the red curve is marked on the plot with a circle.}
\label{fig:NaK_tio}
\end{figure*}

The model spectra and temperature profiles are shown in Fig.~\ref{fig:profiles}. The general behaviour of the temperature profiles are consistent with expectations for strongly irradiated H$_2$-rich atmospheres, as also reported in previous studies \citep[e.g.,][]{burrows_2008,molliere_code}. An isotherm occurs for all of these systems at pressures above $\sim$1 bar. Below the isothermal layer is the convective zone, the transition to which is marked with a circle for each P-T profile in Fig.~\ref{fig:profiles}. The convective energy transport is due to the heat emanating from the planetary interior, taken to be at a temperature of $75K$ for this demonstration. Above the isothermal layer the temperature decreases monotonically outward and ultimately approaches isotherms again in the very low optical depth regime, and as expected, the temperature profiles become consistently hotter with increasing irradiation. The equilibrium chemistry beyond $\approx1500$ K is similar, dominated by CO and H$_2$O (see Fig. \ref{fig:eqm_chem}) due to which the profiles do not show much variation in the gradient. The atmosphere is entirely radiative down to \textasciitilde$10^3$ bar, with the radiative-convective boundary appearing in the region marked with a circle. This can only be seen for the lowest equilibrium temperature of $1000$ K. Higher stellar fluxes increase the depth of the radiative zone owing to the strength of the incoming radiation, driving the convection zone deeper into the atmosphere. Eventually all of the P-T profiles converge onto the same adiabat at great depth, which is determined most strongly by the internal heat flux and the opacity. Lack of strong visible opacity can also be responsible for suppressing the convective regions to lower in the atmosphere (for instance see Fig. \ref{fig:NaK_tio}), and this is discussed in more detail in \ref{sec:visible}.

The planet-star flux ratios reveal the interplay between the temperature profiles and chemistry. A negative temperature gradient in the temperature profile results in strong absorption features in the emergent spectrum as shown in Fig.~\ref{fig:profiles}. Again, the general behaviour of these spectra is as expected and extensively discussed in the literature \citep[see e.g.,][]{Madhu_inversions2010,burrows_2008}. The stellar spectrum peaks in the visible whereas the planetary spectra for these temperatures peak in the near-infrared, implying that the planet-star flux ratio increases with wavelength in the infrared until eventually becoming constant in the mid-far infrared in the Rayleigh-Jeans limit. Naturally, the spectrum is brighter for hotter planets. The locations and amplitudes of the features in the spectra are driven by the temperature gradient and the abundances of absorbing molecules. As alluded to above, the temperature profiles with negative gradients give rise to the observed absorption features at wavelengths where the the abundance molecules absorb. The peaks in the spectrum probe deeper regions of the atmosphere where as the troughs probe regions higher up. In the current temperature range for H$_2$-rich atmospheres in chemical equilibrium, the prominent molecular opacity is provided by H$_2$O in most of the observed features, followed by CO in the 4-6 $\mu$m region. Additional, features can be contributed by other species such as CH$_4$ and CO$_2$ depending on the temperature, metallicity, C/O ratio, etc.  While the troughs in the spectra correspond to the absorption features, the peaks correspond to windows in molecular opacity where the temperature in the isothermal deep atmosphere is probed (deeper than \textasciitilde$1$ bar). On the other hand, the presence of a thermal inversion reverses this symmetry whereby the peaks would correspond to molecular emission features, as discussed in section ~\ref{sec:visible}, and also discussed in previous works \citep[e.g.,][]{Madhu_inversions2010,burrows_2008}.

\subsubsection{Effect of Metallicity and C/O Ratio}
\label{sec:metallicity}

Here we investigate the effect of chemical composition on the model spectra and P-T profiles. We explore the dependence of the spectra on the elemental abundances via two key parameters, the overall metallicity and the C/O ratio, which are expected to govern the dominant molecular composition in hot Jupiter atmospheres \citep[][]{lodders2002,Madhuco_2012,moses2013, heng2015}. The molecular abundances are determined from the elemental abundances of O, C, and N under the assumption of thermochemical equilibrium as discussed in section~\ref{eqm_chem}; the baseline model assumes solar abundances with a C/O = 0.5. In hot Jupiter atmospheres in the 1000-3000 K temperature range the prominent molecules expected are typically H$_2$O, CO, CH$_4$, CO$_2$, and a few other trace species. The relative abundances of these molecules depend on the particular temperature and C/O ratio \citep{Madhuco_2012,moses2013}. For solar abundances, H$_2$O and CH$_4$ are the dominant molecules at T $\lesssim$ 1300 K where as H$_2$O and CO dominate at higher temperatures. Increasing the metallicity of all the elements uniformly, i.e. keeping the C/O constant, increases all these molecular abundances almost linearly but with slightly higher enhancement of CO$_2$ \citep{madhu_seager_2011}. As the C/O ratio is increased the carbon-based molecules become more abundant at all temperatures. A particularly non-linear effect arises at high temperatures ($\gtrsim$1300 K) for which as the C/O approaches 1 the H$_2$O abundance drops substantially. Therefore, the C/O ratio is expected to have a particularly high effect on high-temperature atmospheres  \citep{Madhuco_2012}. Therefore, in what follows we assess the effect of metallicity and C/O on the models for two different temperatures, 1000 K and 1500 K. 

We discuss here models for three representative metallicities of solar, $10\times$ solar, and $30\times$ solar, assuming a C/O ratio of 0.5. And, in each case we investigate two representative temperatures of 1000 K and 1500 K. The results are shown in Fig.~\ref{fig:metallicity}. Increasing the metallicity uniformly across all the elements effectively amounts to a nearly linear increase in the opacity via the increased molecular mixing ratios. For a given irradiation, or $T_{\rm eq}$, the higher opacity increases the atmospheric absorption and causes the temperature profiles to be systematically hotter and their gradients to be steeper. Additionally, due to the increased opacity the onset of the isotherm in the lower atmosphere also happens earlier (i.e. at lower pressures or higher altitudes). All these effects are even stronger for higher $T_{\rm eq}$. The higher temperatures also suppress convective regions deeper in the atmosphere, and so the atmospheres are mostly radiative at the pressures modelled (only the solar $1000$K P-T profile shows the radiative-convective boundary in Fig. \ref{fig:metallicity}). The consequent effect on the spectra is deeper spectral features for higher metallicities caused by not only the hotter and steeper P-T profiles but also the increased abundances of the molecules causing the absorption. The differences in the spectra between the $T_{\rm eq}$ of 1000 K and 1500 K is due to the differences in molecular abundances as a function of temperature, with CO being more abundant in the higher temperature case causing the features in the 4-6 $\mu$m region, whereas CH$_4$ and CO$_2$ being more abundant in the lower temperature case causing the spectral features in the 3-5 $\mu$m and $\sim$15 $\mu$m regions. Overall, however, the effect on the spectra over this metallicity range is relatively modest \citep[also see][]{molliere_code}, as the H$_2$O absorption troughs which dominate the spectra are nearly saturated. 

The spectra are more strongly affected by changes in the C/O ratio than by the overall metallicity, particularly for high temperature atmospheres. As discussed above, the changes in C/O ratio mainly influence the composition by increasing the C-based molecules relative to the O-based molecules. The effect is only marginal for $T_{\rm eq}$ of 1000 K as at such low temperatures most of the C is in CH$_4$ and most of the O is in H$_2$O, irrespective of the C/O ratio. On the other hand, for higher temperatures the H$_2$O abundances can be over 100$\times$ lower for C/O $\gtrsim$ 1 relative to solar C/O ratio. The corresponding effect on the spectrum is quite clear, as shown in Fig. \ref{fig:co_ratio}. In the $T_{\rm eq}$ = 1500 K case, it is important to note the substantial differences in the spectra despite the marginal differences in the temperature profiles, which indicates that the differences are predominantly due to the change in chemistry. The temperature profile for the C/O = 0.5 case is marginally hotter than the high C/O cases because of higher opacity in the former due to the high H$_2$O opacity; this effect is similar to the high metallicity cases discussed above. The differences in the spectra are most prominent in the H$_2$O absorption bands across the spectral range. 

\begin{figure*}
 \includegraphics[trim = 11mm 0mm 4mm 0mm, clip,width=\textwidth]{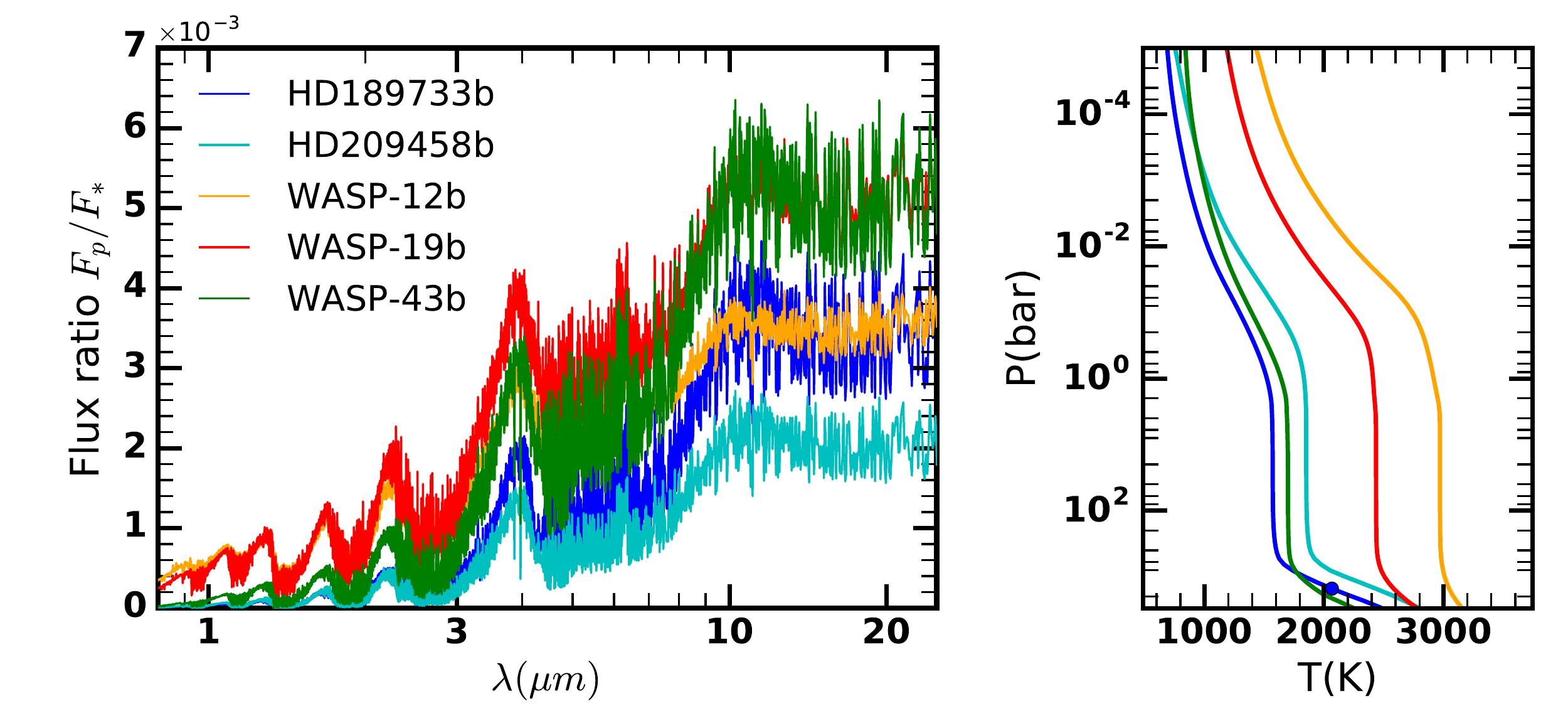}
 \caption{Model spectra and P-T profiles for several known hot Jupiters. The planetary and stellar parameters are given in Table \ref{table:planet_specs}. The models are discussed in section~\ref{sec:known_systems}.}
\label{fig:planet_fluxes}
\end{figure*}

\subsubsection{Visible Absorbers and Thermal Inversions}
\label{sec:visible}

Here we investigate the effect of visible opacity on the spectra and P-T profiles. In the previous sections we investigated the effect of infrared opacity due to prominent C and O molecules on the atmospheric structure and spectra. However, several species are expected to be prevalent in hot Jupiter atmospheres that can provide strong visible opacity, e.g. the alkali atoms Na and K, and metal oxides (TiO, VO, etc.). While the Na and K absorption is primarily due to strong doublet resonance lines centred at $\sim$0.59$\mu$m and $\sim$0.77$\mu$m, respectively, and appropriately broadened, the absorption due to TiO, VO, etc., are very strong and broad bands spanning almost the entire optical spectrum and hence contributing much stronger opacity. On the other hand, TiO is not expected to be as abundant in gas phase as Na/K due to the significantly lower abundance of Ti, much higher condensation temperature, and much more prone to settling and C/O ratios \citep[e.g.][]{spiegel2009,knutson2010,madhu_2011_nature}. Nevertheless, TiO has been suggested as a potential candidate to cause thermal inversions in the hottest of hot Jupiters \citep{hubeny_2003_tio,fortney_2008} and has also been suggested in 2 systems \citep{haynes2015,evans2016}. Therefore, we explore the dependence of spectra and P-T profiles on both Na/K and TiO. 

Considering Na and K in the atmosphere \citep{burrows_sodium} at solar abundances (at pressures greater than 0.1 bar) increases the opacity in the visible part of the spectrum, near the peak of the stellar flux. The left panel in Fig.~\ref{fig:NaK_tio} shows the comparison between models with and without Na/K absorption. The increased visible opacity leads to greater absorption of the incident star light in the upper layers of the atmosphere making the upper temperature profile hotter. It also means that less of the stellar flux penetrates deeper down thereby causing the lower isothermal layer to start higher up in the atmosphere and so does the radiative-convective boundary. Even small quantities of sodium and potassium at solar abundances are adequate to significantly influence the P-T profile, as  shown in Fig.~\ref{fig:NaK_tio}, and the resultant flux is substantially increased as well. 

\begin{figure}
 \includegraphics[trim = 14mm 0mm 24mm 0mm, clip,width=\columnwidth]{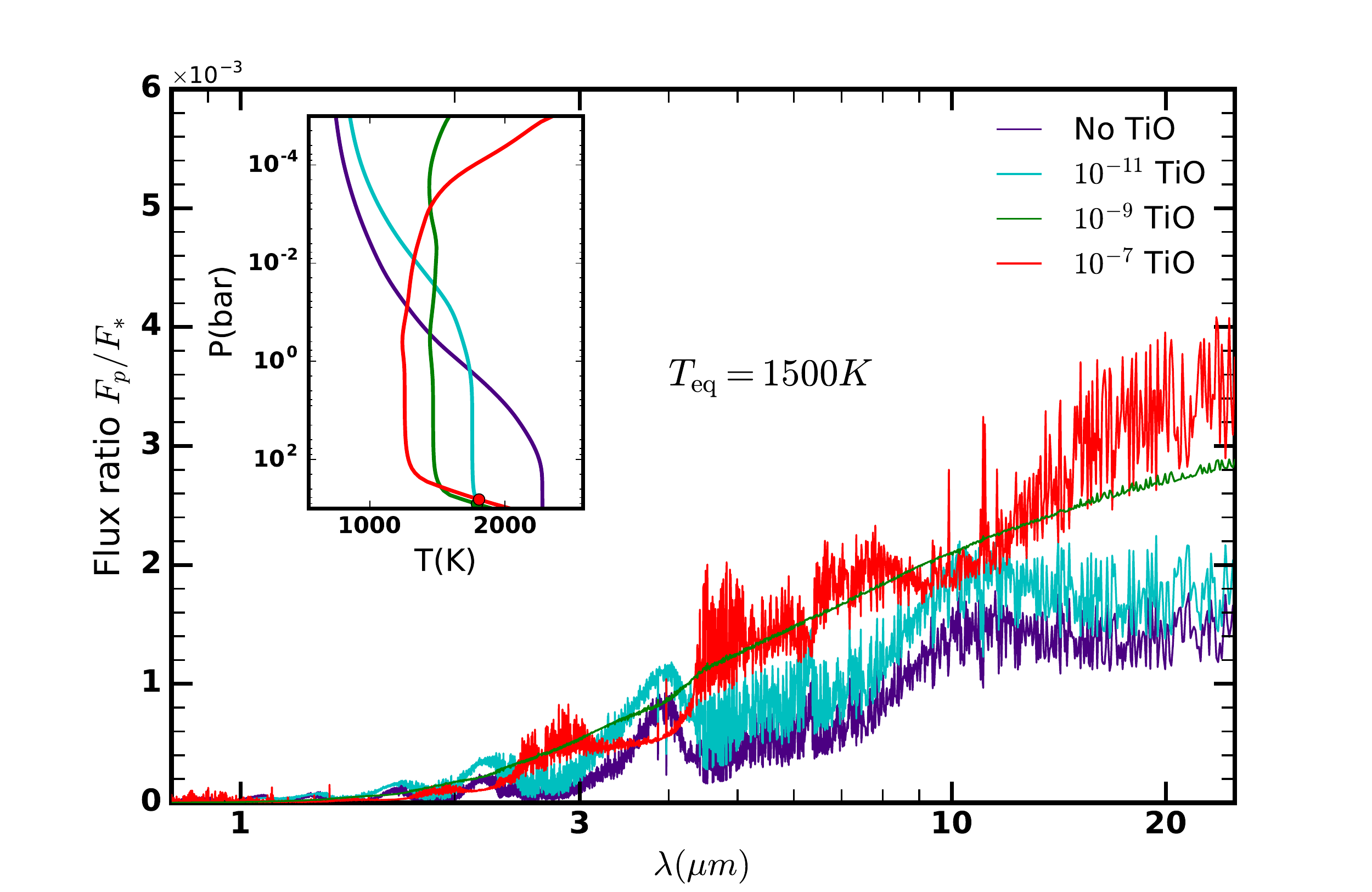}
 \caption{Effect of TiO absorption on the spectrum and P-T profile of a hot Jupiter. A model with equilibrium temperature of $1500$ K is considered with varying levels of TiO in the atmosphere with the mixing ratio ranging from zero to $10^{-7}$.}
\label{fig:tio}
\end{figure}

Now we consider a model planet which has only TiO as the visible absorber. As it is only present as a gas at the very highest temperatures that exoplanets can be, we decided to model WASP-12b, with an equilibrium temperature of almost $3000$ K (see table. \ref{table:planet_specs}). If TiO is present at solar abundance instead of sodium or potassium (see fig. \ref{fig:NaK_tio}), a temperature inversion occurs at the top of the atmosphere and totally transforms the P-T profile and flux ratio. The very strong TiO opacity means more of the stellar flux is absorbed near the top of the atmosphere. This causes the top of the atmosphere to heat up, and keeps the lower layers cooler by preventing the peak stellar flux penetrating down. The increase in opacity leads the convective region to move up higher into the atmosphere as well. TiO is very effective at absorbing the incident stellar flux in the visible, and is able to influence the atmospheric temperature profile significantly despite being present in much smaller quantities than other prominent volatiles species \citep{fortney_2008,Madhuco_2012}. The inversion created means that absorption features in the infrared now become emission features, coming from the hotter regions higher up. An inverted profile is an indication of very strong opacity in the visible near the top of the atmosphere, and indeed only a few data points are needed to verify this \citep{Diamond-Lowe_2014,Madhu_inversions2010}. 

The effect of TiO on the temperature profile and spectrum can be seen in Fig. \ref{fig:tio}. As the TiO abundance is gradually increased the P-T profile transitions from having no inversion to having a strong inversion, for a nominal hot Jupiter with an equilibrium temperature of $1500$ K. The upper layers become hotter as more TiO is added, and the lower layers have a reduced flux incident upon them and hence are cooler. The green line is the transition point at which absorption features become emission features, resulting from the inversion.

\subsection{Models of Known Irradiated Planets - Effect of System Properties}
\label{sec:known_systems}
We use GENESIS to generate models of several known hot Jupiters to investigate the effect of system parameters, particularly the stellar properties. We consider five well studied hot Jupiters spanning a wide range in temperature ($T_{\rm eq}$$\sim$1200 - 2600 K): HD~189733b (1200 K), WASP-43b (1375 K), HD~209458b (1450 K), WASP-19b (2065 K), and WASP-12b (2580 K). The system parameters are shown in Table~\ref{table:planet_specs} and the model spectra and temperature profiles are shown in Fig.~\ref{fig:planet_fluxes}. We generate fiducial models assuming solar values for the elemental abundances, and consider only the prominent O, C, and N based molecular species. We also include Na and K absorption and collision induced opacity due to H$_2$-H$_2$ and H$_2$-He. In each case, the stellar flux was modelled using the Kurucz spectrum for the corresponding stellar properties shown in Table~\ref{table:planet_specs}.

As expected, both the incident irradiation at the planet and the stellar spectrum have a palpable effect on the atmospheric temperature profile. A greater irradiation, represented by $T_{\rm eq}$, causes a hotter temperature profile for the same chemical composition, as can be seen in Fig.~\ref{fig:planet_fluxes}. Consequently, the P-T profiles of HD~189733b and WASP-12b are the coolest and hottest in our sample. On the other hand, despite the $T_{\rm eq}$ for WASP-43b and HD~209458b being similar, the P-T profiles have different gradients. This is due to the difference in the stellar spectra. WASP-43 is a cooler star (see table \ref{table:planet_specs}), implying that the peak of its spectrum is at a longer wavelength, due to which the incident radiation penetrates deeper down causing a steeper temperature gradient. These models, along with the results in section~\ref{sec:visible}, demonstrate that the temperature gradient in a highly irradiated atmosphere is influenced by several factors including the incident irradiation, the stellar spectrum, and the chemical composition, both of visible and infrared absorbers. 

\begin{table*}
\begin{tabular}[t]{l*{8}{c}c}
Planet              & $Z_{\mathrm{star}}$ & $T_{\mathrm{eff}}$& $R_{\mathrm{star}}$ & $\mathrm{log}(g_{\mathrm{star}})$ & $a$  & $T_{\mathrm{eq}}$  & $R_p$ & $\mathrm{log}(g_{\mathrm{planet}})$ & Distance to system \\
              & & (K) & $(R_{\mathrm{sun}})$ & (c.g.s) & (A.U.) & $(K)$  & ($R_J$)& (c.g.s) & (pc) \\
\hline
HD189733b & -0.03 & 5040 & 0.756 &4.587 & 0.0310 & 1200 & 1.14 & 3.34 & 19.5  \\
HD209458b  & 0.00 & 6065 & 1.155 &4.361 & 0.0472 & 1450 & 1.359 & 2.97 & 49.6  \\
WASP-12b   & 0.30 & 6300 & 1.63 &4.38 & 0.0225 & 2580 & 1.79 & 2.99 & 267.0  \\
WASP-19b    & 0.02 & 5500 & 0.99 &4.5 & 0.0163 & 2065 & 1.39 & 3.17 & 250.0  \\
WASP-43b    & -0.05 & 4400 & 0.598 &4.65 & 0.0142 & 1375 & 0.930 & 3.71 & 80.0  \\
\end{tabular}
\caption{System parameters of known hot Jupiters modelled in section~\ref{sec:known_systems} and Fig. \ref{fig:planet_fluxes}. $Z_{\rm star}$ is the stellar metallicity, $T_{\rm eff}$ is the effective temperature of the star and $T_{\rm eq}$ is the equilibrium temperature of the planet. The equilibrium temperature is calculated assuming the albedo is 0, and complete redistribution of flux over the planet, $T_{\mathrm{eq}} = T_{\mathrm{eff}}~\sqrt[]{R_{\mathrm{star}}/(2a)}$. The system parameters are obtained from exoplanets.org.}
\label{table:planet_specs}
\end{table*}

The stellar properties also have a significant effect on the observed planet-star flux ratio. Generally, for a given host star, hotter and larger planets have higher emergent fluxes. However, for transiting exoplanets the key observable quantity is the planet-star flux ratio. This implies for the same planetary properties a cooler and/or smaller star would lead to larger planet-star flux ratio. Fig.~\ref{fig:planet_fluxes} demonstrates this natural expectation. At the cooler end, the planets HD~189733b, WASP-43b, and HD~209458b have similar equilibrium temperatures (1200-1450 K), with HD~209458b being the hottest. However, WASP-43b has over twice the planet-star flux ratio compared to HD~209458b, with HD~189733b being in between, owing to the the host star WASP-43 being the coolest star in our sample. Similarly, at the hotter end WASP-19b and WASP-12b have similar $T_{\rm eq}$ but the flux ratio for WASP-19b is significantly higher than WASP-12b, again owing to its smaller and cooler star. As an extreme case, WASP-12b has nearly twice the $T_{\rm eq}$ of WASP-43b but still lower flux ratio than the latter. Thus, the stellar parameters play a key role in determining the observability of emergent spectra of exoplanets, which justifies the numerous current exoplanet searches around cooler and smaller stars. 

\subsection{Non-irradiated Atmospheres}

GENESIS has the capability to simultaneously and self-consistently consider both external stellar irradiation and internal flux from within the planet. We now turn to models of non-irradiated atmospheres which are relevant for planets on large orbital separations or free floating planets and brown dwarfs. Such models are particularly useful for directly-imaged sub-stellar objects, young giant exoplanets \citep{marois2010,bonnefoy2014,barman2015,macintosh2015} and brown dwarfs \citep{burgasser_spex2014,apai2013} detected via direct imaging and for which high quality spectra are becoming available. Several conventional models of non-irradiated objects exist in the literature \citep[e.g.][]{ marley2012,madhusudhan2011_hr8799,barman2015} and include cloud-free as well as cloudy models, unlike self-consistent models of irradiated hot Jupiters which are generally cloud free \citep[e.g.][]{burrows_2008,fortney_2008}. In the present work, we explore only cloud-free models of non-irradiated objects and will consider cloudy models in future work. 

For our present exploration, we will again model a Jovian-like planet but now with an increased internal temperature that is much greater than the 75 K we considered for irradiated planets; $T_{\mathrm{int}}$ is now $1500$ K unless otherwise specified and the stellar irradiation is negligible. We nominally consider a solar-type star with the orbital separation of the planet at 10 AU. In direct imaging the observable is the emergent spectrum directly from the planet and not the planet-star flux ratio. We assume the distance to the system to be 10 parsecs. The chemical composition, again, is determined via chemical equilibrium with solar elemental abundances. In what follows, we first compare the differences in the emergent spectrum and P-T profile for irradiated and non-irradiated planets with matching equilibrium and internal temperatures. We then explore how the internal flux and metallicity influences the profile and spectrum for non-irradiated planets.  

\subsubsection{Non-irradiated vs. Irradiated Planets}
\label{sec:irr_noirr}
\begin{figure}
 \includegraphics[trim = 4mm 0mm 24mm 0mm, clip,width=\columnwidth]{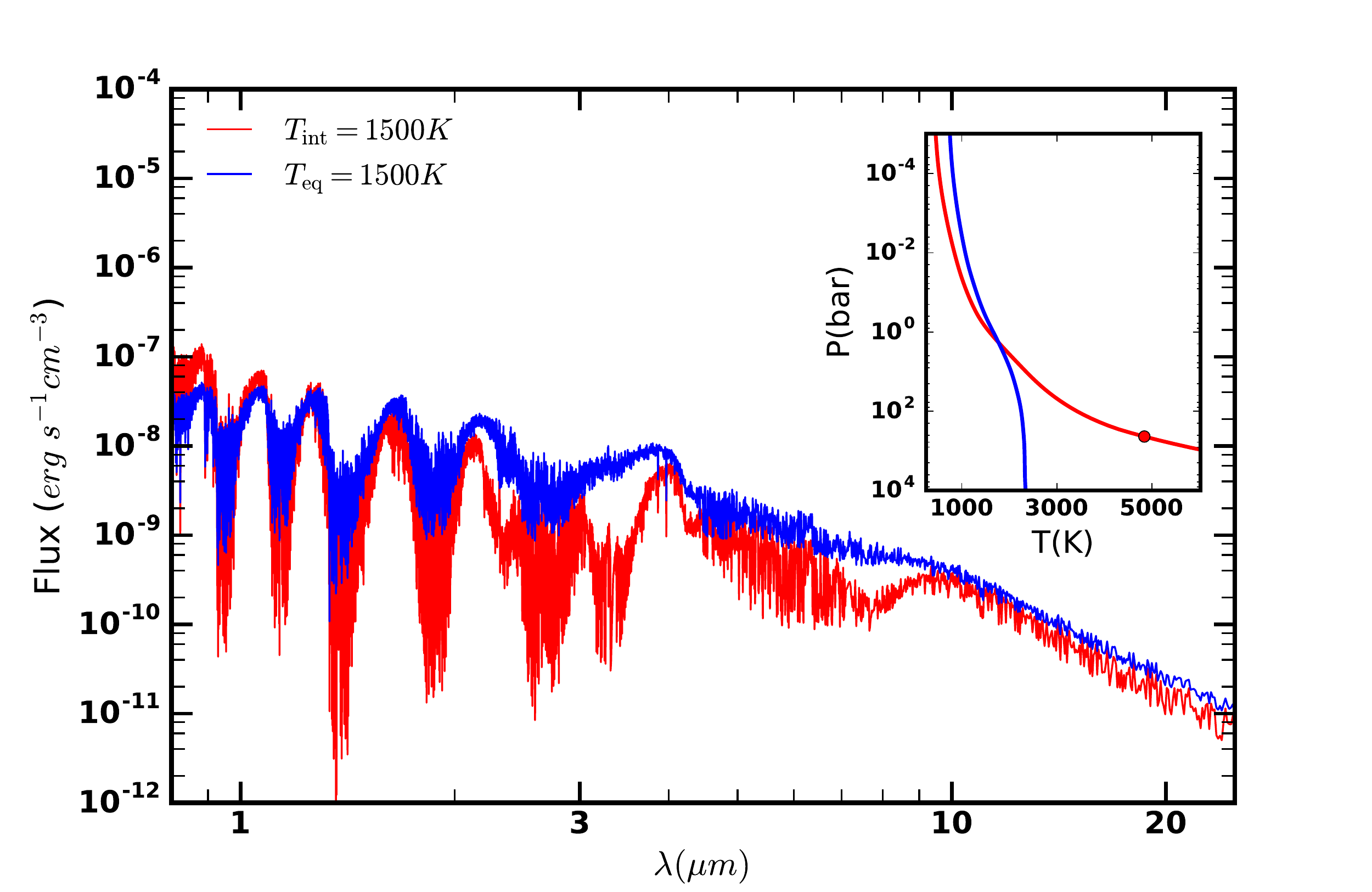}
 \caption{Comparison between an irradiated and a non-irradiated giant planet. The two models show the emergent spectra and P-T profiles for an irradiated hot Jupiter with an equilibrium temperature of 1500 K (blue) and a non-irradiated Jupiter with an internal temperature of 1500 K (red), as discussed in section~\ref{sec:irr_noirr}. Both the planets are assumed to be 10 pc away from the observer. The circle in the P-T profile indicates the radiative-convective boundary.}
\label{fig:direct}
\end{figure}

The differences in the observed spectra for planets with strong external versus internal flux are shown in Fig. \ref{fig:direct}. The fluxes set the top/bottom boundary conditions for the irradiated and non-irradiated atmospheres, respectively. For the irradiated planet we consider a $T_{\mathrm{eq}}$ =1500 K and $T_{\mathrm{int}}$ = 75 K whereas for the non-irradiated planet we consider negligible irradiated and $T_{\mathrm{int}}$ = 1500 K. The strong external flux applied at the top of the atmosphere for the irradiated planet leads to a hotter region near the top. However, the internal flux for the non-irradiated planet causes the deep layer temperature to be hotter. Strong external irradiation suppresses the convective region of the atmosphere to deeper in the atmosphere, below the pressure range in the inset figure, and the temperature tends to an isotherm in the lower atmosphere as seen in the previous sections. On the other hand, the radiative-convective boundary can be seen for the non-irradiated planet. 

The spectrum has greater emission for the irradiated planet in the infrared, as the temperature in the regions where the emission occurs is greater, i.e. higher up in the atmosphere above the $\sim$1 bar level. The non-irradiated planet has the higher emission at $\sim$$1\micron$ where there is little opacity at these wavelengths and hence deeper regions below the $\sim$1 bar level are probed, where the internal flux dominates. The absorption features for the non-irradiated planet are also much more pronounced, and show greater flux differences. This is owing to the greater temperature gradient for the non-irradiated atmosphere. Fig. \ref{fig:direct} thus shows a clear difference between the spectra and P-T profiles of highly irradiated planets observed using transit spectroscopy and weakly irradiated planets observed by direct imaging. 

\subsubsection{Effect of Internal Flux}

\begin{figure}
 \includegraphics[trim = 4mm 0mm 24mm 0mm, clip,width=\columnwidth]{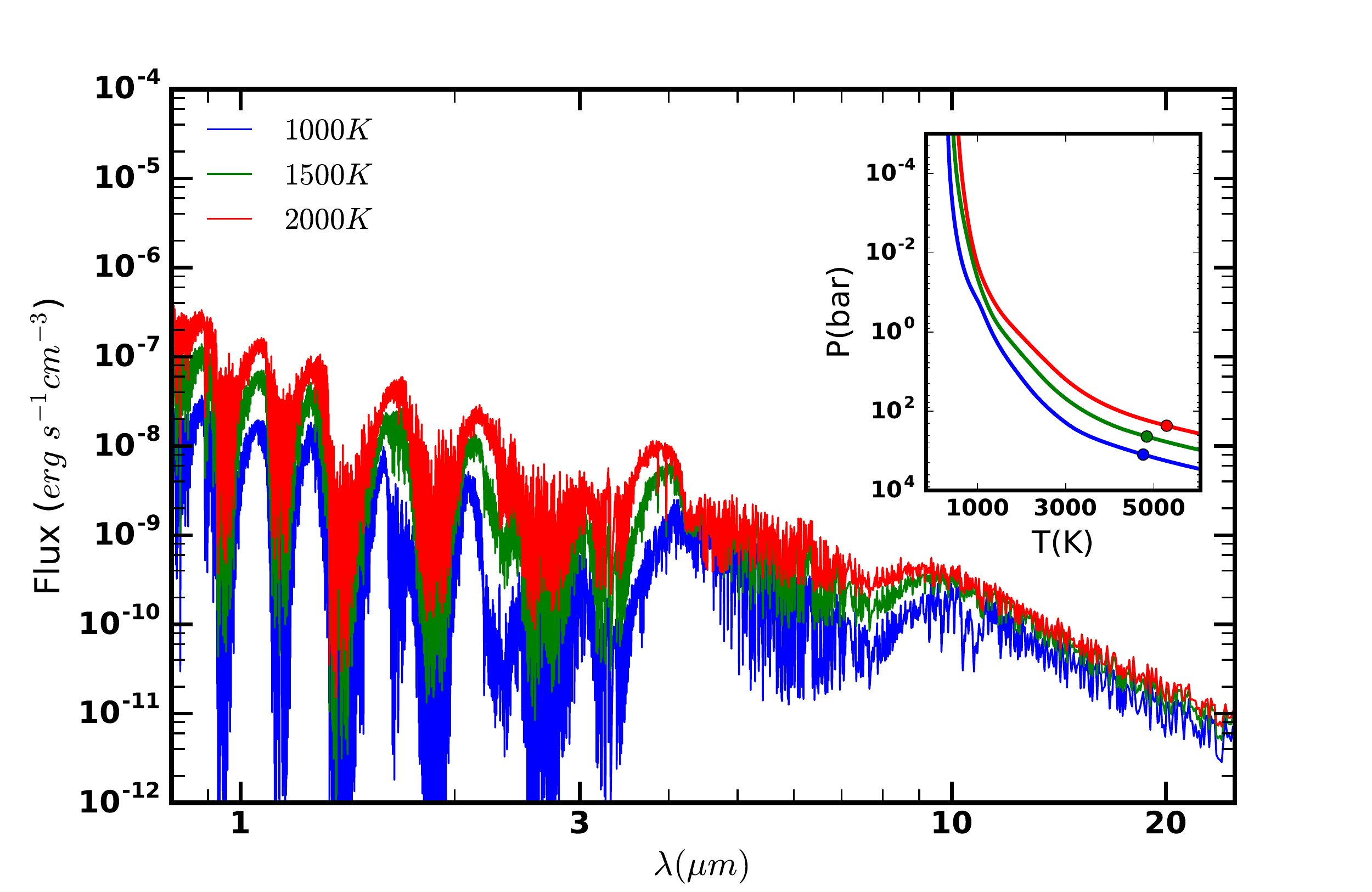}
 \caption{Effect of internal heat on emergent spectra and P-T profiles of non-irradiated planets. A Jupiter placed at 10 A.U. is modelled with different internal heat fluxes represented by $T_{\mathrm{int}}$. The emergent flux is shown for the source at 10 pc from the observer. The corresponding P-T profiles are shown in the inset. The circles denote the radiative-convective boundary.}
\label{fig:Tint}
\end{figure}

We explore models of our fiducial non-irradiated planet described above with the internal flux, represented by an effective internal temperature $T_{\mathrm{int}}$, varied between $1000$ K to $2000$ K. The elemental abundances are kept fixed at solar values and thermochemical equilibrium is assumed; the molecular composition however may change depending on the temperature structure. The emergent spectra and P-T profiles are shown in Fig. \ref{fig:Tint}. Naturally, higher internal flux leads to hotter temperature profiles in the atmosphere. In particular the deeper regions are warmer as higher flux is transported upwards from within the planet. The radiative-convective boundary is also shifted upwards and to a higher temperature as well; the increase in internal flux drives a greater convective flux. Considering the lowest pressures, the external irradiation is now negligible, and so the atmospheres all tend to the same isotherm, one where the escaping heat balances the heat from the core. The internal heat has a very weak influence on the top layers of the atmosphere, as the optical depth and hence absorption of the flux is low. There are slight differences in the P-T gradient arising from varying chemistry, and vice versa, particularly at $\sim1000$ K where the abundances (therefore the optical depth and absorption) are the most sensitive to temperature. The emergent spectra show greater emission for the planets with greater internal temperature for every wavelength, perhaps unsurprisingly as every point of the atmosphere is hotter. The figure shows clearly that the internal heat flux, although often considered deep inside the planet, strongly influences the observed spectrum, with amplitudes of spectral features of even several orders of magnitude when considering some of the absorption features.  

\begin{figure*}
 \includegraphics[trim = 33mm 0mm 42mm 0mm, clip,width=\textwidth]{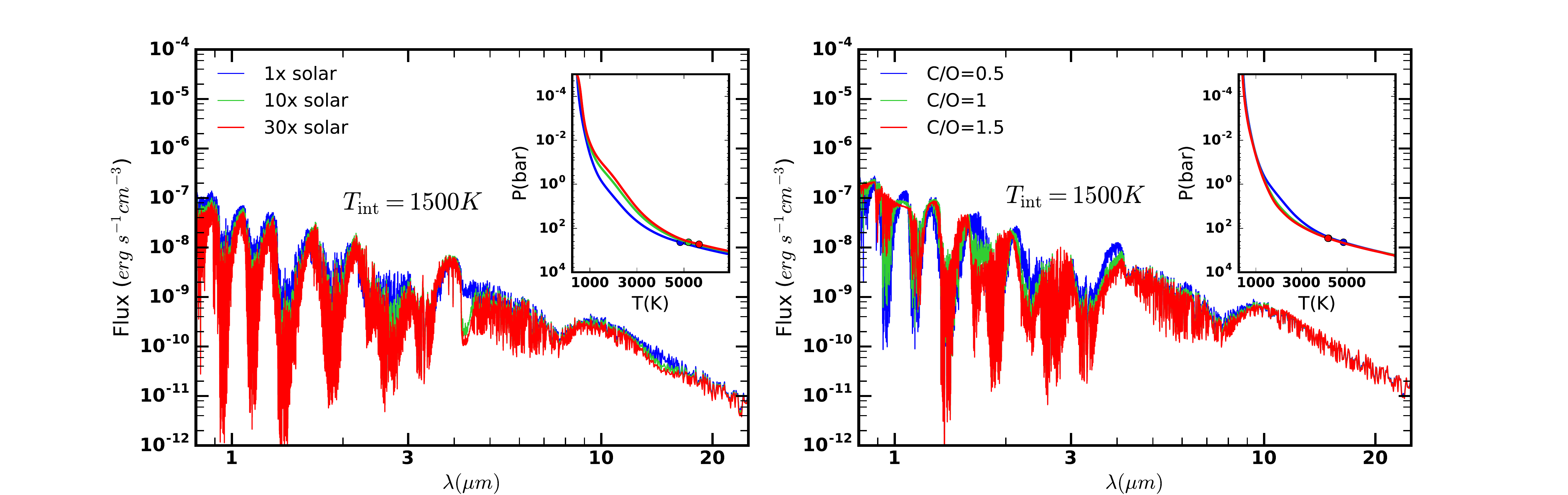}
 \caption{Effect of metallicity and C/O ratio on emergent spectra and P-T profiles for non-irradiated planets. A jupiter-sized planet with an internal temperature of 1500 K is considered. The metallicity is explored between 1-30 $\times$solar and the C/O  ratio spans 0.5-1.5.}
\label{fig:Tint_met}
\end{figure*}

\subsubsection{Effect of Metallicity and C/O ratio}

We also report models over a range of metallicities, as shown in Fig.~\ref{fig:Tint_met}. The metallicities of C, O, and N considered were solar, 10$\times$ solar, and 30$\times$ solar, with the internal temperature now fixed at $1500$ K, analogous to Fig.~\ref{fig:metallicity} for irradiated planets. Note that now the resultant P-T profile is not as different, with only a slightly greater temperature seen for the highest metallicities. The radiative-convective boundary is also not dissimilar, occurring near similar temperatures (\textasciitilde$5000$ K) and pressures (\textasciitilde$1000$ bar) for each case. 

The flux spectrum does show nominal differences between the different metallicities, with the higher flux observed for the solar metallicity due to lower absorption. The strong opacity for the green and red curves mean the emission occurs from lower pressures, where it is cooler. The features in the flux plot are more pronounced for the highest metallicity due to the greater molecular abundance leading to greater absorption in the molecular bands. Furthermore, the strong feature in the 4-5 $\mu$m region in the higher metallicity and higher temperature models is due to the higher abundance of CO and CO$_2$. Altering the C/O ratio does alter the received flux more significantly, as increased C/O ratios beyond 1 reduce the water abundance by orders of magnitude (especially at temperatures exceeding $1000$ K, here the oxygen is taken up in the CO). This reduces the spectral features of H$_2$O, and hence even though the P-T profile is largely unaffected, the chemistry and therefore the flux has observable differences.

We have explored models of non-irradiated planets over a range in metallicity, C/O ratio and internal flux. The P-T profiles and emergent spectra show clear differences compared to those of irradiated planets, as expected and found in previous works. In the future, more models exploring the parameter space in greater detail would be beneficial, particularly with the inclusion of clouds that are inferred for several substellar objects \citep{marley2012,madhusudhan2011_hr8799,barman2015} and brown dwarfs \citep{burgasser_spex2014,apai2013}. 

\section{Summary and Discussion}
\label{sec:summary}
We introduce a new code GENESIS for modelling exoplanetary atmospheres. Our effort is driven by current and upcoming high-resolution and high-precision spectra of exoplanets and brown dwarfs. Current observations of exoplanetary spectra span a variety of methods, ranging from transit spectroscopy and direct imaging to very high resolution Doppler spectroscopy with $R\sim10^5$. On the other hand, observational inferences and theoretical studies are suggesting that exoplanetary and brown dwarf atmospheres can span a wide range of physicochemical conditions, e.g., incident irradiation, metallicities, C/O ratios, and internal fluxes. Consequently, there is a growing need in the field for high-fidelity self-consistent models of exoplanetary spectra that span over all these parameters and a wide range in spectral resolution. Several self-consistent models exist in the field, mostly derived from pre-existing codes e.g., from stellar or solar-system applications, with varied degrees of complexity and applicability to these requirements but also suggesting the need for a new generation of models, as  discussed in section~\ref{sec:introduction}. Our present work is a step in that direction. 

The GENESIS code is custom-built for self-consistent, line-by-line, plane-parallel modelling of exoplanetary atmospheres with a high degree of robustness and applicability. The code can be used to model exoplanetary atmospheres over a wide range in incident irradiation, metallicity, C/O ratio, internal fluxes, etc., and can be used to model irradiated as well as non-irradiated atmospheres. The code uses well-tested robust and accurate methods for each component of the model. The radiative transfer is solved line-by-line using the Feautrier method which allows for treatment of full angular dependance of the intensity and is second order accurate. The code ensures radiative-convective equilibrium using the Rybicki's method with complete linearisation to find the temperature correction. The radiative and convective regions of the atmosphere are solved simultaneously and self-consistently, treating the external irradiation and internal flux as the top and bottom boundary conditions, respectively. The line-by-line absorption cross sections of the chemical species are derived from the latest high temperature line lists, including the effect of thermal and pressure broadening wherever available. The atmosphere is assumed to be in chemical equilibrium with the temperature profile and, given the elemental abundances, the main constituent gas mixing fractions are found using established methods. 

In the present work, we demonstrate our code for modelling giant exoplanetary atmospheres, both irradiated and non-irradiated, which are currently the most studied objects in the field. The code is tested extensively for consistency and robustness and against published models of some known hot Jupiters generated using existing codes. We generate grids of models in the parameter space for both irradiated and non-irradiated objects. For irradiated planets, we explore the effects of $T_\mathrm{eq}$, metallicity and C/O ratio on the P-T profile and emergent spectrum, and find trends consistent with previous studies. We note how the C/O ratio has a significant effect on the spectrum, particularly at high temperatures. We also discuss the effects of visible opacity, e.g. of Na/K and TiO, and, particularly, the formation of thermal inversions in hot Jupiters atmospheres as a function of TiO abundance. Finally, we investigate model of several known hot Jupiters are discuss the effect of the stellar properties on the observable atmospheric properties. We also investigate models for non-irradiated planets observing how the P-T profile and emergent flux vary with the internal flux and metallicity, and demonstrate the significant differences between the spectra and P-T profiles of irradiated and non-irradiated atmospheres. This provides a good demonstration of the robustness and diversity of the model in its ability to model across the parameter space accurately.  

In the present code we make some basic assumptions that are common to all self-consistent equilibrium models of exoplanetary atmospheres. The main basis of the model is a plane-parallel geometry, though our radiative transfer scheme does take into account the angular dependence of the specific intensity in each layer of the atmosphere. We assume general equilibrium conditions, i.e., local thermodynamic equilibrium (LTE), hydrostatic equilibrium, radiative-convective equilibrium, and chemical equilibrium. Even though the code is generic for any chemical composition, in the present work we focus on giant exoplanetary atmospheres. As such, we assume the atmosphere to be H$_2$-rich. Furthermore, here we only focus on the dominant chemical compositions expected in such atmospheres, namely considering only the prominent molecular species with C-O-N chemistry in chemical equilibrium. However, the extension to include other molecules is straightforward using a full equilibrium chemistry code \citep[e.g.][]{seager_2005, Madhuco_2012}. We also consider additional species such as Na/K and TiO to demonstrate some cases with strong visible opacity. We have assumed that the scattering is only due to H$_2$ Rayleigh scattering and that the atmosphere is cloud-free. Any non-equilibrium phenomena, e.g. photochemical or dynamical effects, have been ignored.   

The code can be further developed to incorporate new processes that depart from the present equilibrium assumptions. While our current cloud-free models may be applicable to a range of irradiated hot Jupiters as well as to some brown dwarfs (e.g. T Dwarfs), observations and theoretical studies suggest that clouds/hazes can exist in several of these atmospheres \citep{Helling_2008_bd_clouds,marley2012,madhusudhan2011_hr8799,barman2015}. Therefore, we anticipate including clouds/hazes in our model in the future. While some previous models have pursued it in the two-stream approximation \citep{marley2012}, in our case the radiative transfer solver which currently uses the scalar Feautrier method would need to be modified to handle strong scattering, e.g., using either the matrix Feautrier method or the Discontinuous Finite Elements (DFE) method with Accelerated Lambda Iteration \citep[][]{Sudarsky_2003,hubenybook}. Additionally, our model currently does not include any prescription to self-consistently transport energy to the night side, e.g., due to winds, etc. We do indirectly account for that effect by adjusting the incident stellar flux with a parametric factor ($f_\mathrm{star}$), as also pursued by \citep{fortney_2008}. This can be improved in the future by adding an energy sink on the dayside and self-consistently ensuring radiative equilibrium \citep[e.g.,][]{burrows_2008}. We could also incorporate prescriptions in our code to consider non-equilibrium chemical processes. Finally, the chemical and opacity database used in the present model can be expanded significantly beyond the species currently considered. The models can also be easily extended to considering lower mass planets with significantly different compositions from the H$_2$-rich atmospheres explored in our current work.  

The GENESIS models from the present work are aimed to be a valuable resource to the community\footnote{The models are made publicly available at \url{https://github.com/exo-worlds/genesis}}. We provide self-consistent models of emergent spectra and pressure-temperature-density profiles over a grid in parameter space of giant exoplanetary atmospheres. Additionally, we also provide models of several known hot Jupiters based on their system parameters. In the future our models could be computed at very high resolution. In particular, this is highly desirable for atmospheric detections using high dispersion spectroscopy where accurate high-resolution (R$\sim$$10^5$) models are cross-correlated with an observed spectrum to detect a molecule \citep[e.g.,][]{brogi2012}. Since the models can simulate both irradiated and non-irradiated atmospheres they are applicable to a wide range of objects, spanning the entire range of irradiated hot Jupiters (1000-3000 K) and non-irradiated giant planets and also brown dwarfs. Given its high computational efficiency, our code can also be integrated into radiative transfer components of higher dimensional models. For example, it can be integrated with minimal adaptation into detailed non-equilibrium chemistry codes  \citep[][]{moses2013} to compute the chemical and P-T profiles self-consistently, and similarly into general circulation models \citep[][]{showman_gcm}. As such, the GENESIS models would be valuable both in the planning and interpretation of observations as well as in detailed theoretical understanding of the physical and chemical processes of exoplanetary atmospheres. This is all the more timely given the impending renaissance in atmospheric characterisation of exoplanets with upcoming observations from current and large facilities on the horizon, e.g. HST, VLT, JWST, E-ELT, etc. 

\section*{Acknowledgements}
\addcontentsline{toc}{section}{Acknowledgements}
We would like to thank Dr Ivan Hubeny for helpful feedback 
on the work. SG thanks Adam Jermyn for discussions on the numerical methods, and Ryan MacDonald for helpful discussion  
on the manuscript. SG acknowledges financial support from the Science and Technology Facilities Council (STFC), UK, towards his doctoral programme.




\bibliographystyle{mnras}
\bibliography{references} 




\appendix
\section{Solving the Transfer Equation}\label{appendix:formal_soln}
Discretising equation \ref{eqn:rtj} for a layer of the atmosphere $i$, frequency $k$ and angle $m$ results in

\begin{multline}\label{eqn:formalsoln}
\frac{\mu_{k,m}^2 ~j_{i-1,k,m}}{\Delta \tau_{i-\frac{1}{2},k,m} \Delta \tau_{i,k,m}} - \frac{\mu_{k,m}^2 ~j_{i,k,m}}{\Delta \tau_{i,k,m}}\left(\frac{1}{\Delta \tau_{i-\frac{1}{2},k,m}} + \frac{1}{\Delta \tau_{i+\frac{1}{2},k,m}}\right) \\ + \frac{\mu_{k,m}^2 ~j_{i+1,k,m}}{\Delta \tau_{i+\frac{1}{2},k,m} \Delta \tau_{i,k,m}} = j_{i,k,m} - S_{i,k,m},
\end{multline}
for the transfer equation, with $\mu$ being the cosine of the angle. Given the source function from the previous iteration (or taken to be the planck function for the first iteration) we can calculate the values of $j_{i,k,m}$. Then
\begin{align}
J_{i,k} = \sum_{m} w_m ~j_{i,k,m},
\end{align}
is the mean intensity of radiation for a layer $i$ and frequency $k$, the $m$th angle for some weight function $w$.

The definitions of $\Delta \tau$ are
\begin{align}
\Delta \tau_{i\pm\frac{1}{2},k} \equiv \frac{1}{2g}((\kappa_{i\pm1,k} & + \sigma_{i\pm1,k})/\rho_{i\pm1} + (\kappa_{i,k} + \sigma_{i,k})/\rho_i)|P_{i\pm1}-P_i|~, \\
\Delta \tau_{i,k} & \equiv \Delta \tau_{i+\frac{1}{2},k} + \Delta \tau_{i-\frac{1}{2},k}~,
\end{align}
which come directly from the definition of $d\tau$ and hydrostatic equilibrium.

\section{Computational Method for Linearisation}
To compute the equations of radiative transfer numerically, all of the major equations need to be discretised, as well as constraints and boundary conditions in section \ref{feautrier method}. Doing this for equation \ref{eqn:f_nu} and the corresponding equation for $g \equiv H_{\nu}(0)/K_{\nu}(0)$ gives
\begin{align}
f_{i,k} = \frac{\sum_m w_m \mu^2 ~j_{m,i,k}}{\sum_m w_m ~j_{m,i,k}}~, \\
g_k = \frac{\sum_m w_m \mu ~j_{m,ND,k}}{\sum_m w_m ~j_{m,ND,k}}~.
\end{align}
The weights are denoted by $w$. Discretising equations \ref{eqn:rtJ} and \ref{eqn:reqmint} for $ND$ layers (indexed by $i$) and $NF$ frequencies (indexed by $k$), numbered from 0 at the bottom of the atmosphere, gives
\begin{multline}
\frac{f_{i-1,k}}{\Delta \tau_{i-\frac{1}{2},k} \Delta \tau_{i,k}} J_{i-1,k} - \frac{f_{i,k}}{\Delta \tau_{i,k}}\left(\frac{1}{\Delta \tau_{i-\frac{1}{2},k}} + \frac{1}{\Delta \tau_{i+\frac{1}{2},k}}\right)~J_{i,k} \\+ \frac{f_{i+1,k}}{\Delta \tau_{i+\frac{1}{2},k} \Delta \tau_{i,k}} J_{i+1,k} = \frac{\kappa_{i,k}}{\kappa_{i,k}+\sigma_{i,k}}J_{i,k} - \frac{\kappa_{i,k}}{\kappa_{i,k} + \sigma_{i,k}}B_{i,k},
\end{multline}
for layers $i = 2$,$3$...$ND-1$,
\begin{multline}
\frac{f_{ND-1,k}J_{ND-1,k} - f_{ND,k}J_{ND,k}}{\Delta \tau_{ND-\frac{1}{2},k}} = g_k J_{ND,k} - H_k^{\mathrm{ext}}\\+ \frac{\Delta \tau_{ND-\frac{1}{2},k}}{2}\left(\frac{\kappa_{ND,k}}{\kappa_{ND,k}+\sigma_{ND,k}}J_{ND,k} - \frac{\kappa_{ND,k}}{\kappa_{ND,k} + \sigma_{ND,k}}B_{ND,k}\right),
\end{multline}
for layer $i=ND$ at the top of the atmosphere and
\begin{multline}
\frac{\left(f_{1,k}J_{1,k} - f_{2,k}J_{2,k}\right)}{\Delta \tau_{\frac{3}{2},k}} = \frac{1}{2}(B_{1,k} - J_{1,k}) +\frac{1}{3}\frac{B_{1,k} - B_{2,k}}{\Delta \tau_{\frac{3}{2},k}} \\ - \frac{\Delta \tau_{\frac{3}{2},k}}{2}\left(\frac{\kappa_{1,k}}{\kappa_{1,k}+\sigma_{1,k}}J_{1,k} - \frac{\kappa_{1,k}}{\kappa_{1,k} + \sigma_{1,k}}B_{1,k}\right),
\end{multline}
for the bottom of the atmosphere layer $i=1$. The equations of radiative equilibrium become
\begin{align}
\sum_{k=1}^{NF} w_k \kappa_{i,k}(J_{i,k} - B_{i,k}) & = 0, \\
\sum_{k=1}^{NF} w_k \left(\frac{f_{i,k}J_{i,k} - f_{i+1,k}J_{i+1,k}}{\Delta \tau_{i+\frac{1}{2}}}\right) & = \frac{\sigma_R}{4\pi}T_{\mathrm{int}}^4,
\end{align}
where the quadrature weights are given by $w$ and the last line is only valid for layers up to $i = ND-1$.
With the additional convective flux these become
\begin{align}
\sum_{k=1}^{NF} &w_k \left(\kappa_{i,k}(J_{i,k} - B_{i,k})\right) +\frac{\rho_i g}{4\pi}\frac{F_{\mathrm{conv},i-\frac{1}{2}} - F_{\mathrm{conv},i+\frac{1}{2}}}{\frac{1}{2}(P_{i-1} - P_{i+1})} = 0, \\
\sum_{k=1}^{NF} &w_k \left(\frac{f_{i,k}J_{i,k} - f_{i+1,k}J_{i+1,k}}{\Delta \tau_{i+\frac{1}{2}}}\right) + \frac{F_{\mathrm{conv},i+\frac{1}{2}}}{4\pi}= \frac{\sigma_R}{4\pi}T_{\mathrm{int}}^4.
\end{align}
The $\Delta \tau$ term is linearised by
\begin{align}
\frac{d \Delta \tau_{i\pm \frac{1}{2}}}{dT_i} &= \frac{\Delta \tau_{i\pm \frac{1}{2}}}{\omega_i+\omega_{i\pm1}} \frac{d \omega_i}{dT_i},
\end{align}
where $\omega_i=(\kappa_i+\sigma_i)/\rho_i$.
The $\kappa$ (and $\sigma$) term is linearised by
\begin{align}
\frac{d \kappa_i}{dT_i} &= \frac{\partial \kappa_i}{\partial T_i}+ \frac{\partial \kappa_i}{\partial P_i}\frac{dP_i}{dT_i}.
\end{align}
For an ideal gas, $dP/dT$ is simply $P/T$.


\bsp	
\label{lastpage}
\end{document}